      [1999/12/01 v1.4c Il Nuovo Cimento]
\documentclass{cimento}

\usepackage{graphicx,amssymb,bbm}  
\usepackage{amssymb}  
\def\ket#1{|#1\rangle}
\def\bra#1{\langle#1|}

\def\bracket#1#2#3{\langle #1 | #2 | #3 \rangle}

\title{Entanglement, randomness and chaos}
\author{G.~Benenti\from{cncs}\from{infn}}
\instlist{\inst{cncs} CNISM, CNR-INFM 
\& Center for Nonlinear and Complex systems,
Universit\`{a} degli Studi dell'Insubria, via Valleggio 11,
I-22100 Como, Italy
\inst{infn} Istituto Nazionale di Fisica Nucleare, Sezione
di Milano, via Celoria 16, I-20133 Milano, Italy}
\PACSes{\PACSit{03.67.Mn}{Entanglement production, characterization, 
and manipulation}
\PACSit{03.67.Lx}{Quantum computation}
\PACSit{03.67.-a}{Quantum information}
\PACSit{05.45.Mt}{Quantum chaos; semiclassical methods}
\PACSit{03.65.Yz}{Decoherence; open systems; quantum statistical methods}}
\begin{document}

\maketitle

\begin{abstract}
Entanglement is not only the most intriguing feature of quantum mechanics,
but also a key resource in quantum information science. Entanglement 
is central to many quantum communication protocols, including dense
coding, teleportation and quantum protocols for cryptography. For
quantum algorithms, multipartite (many-qubit) entanglement is necessary
to achieve an exponential speedup over classical computation. The entanglement
content of random pure quantum states is almost maximal; such states find
applications in various quantum information protocols.
The preparation of a random state or, equivalently, the implementation
of a random unitary operator, requires a number of elementary one- and
two-qubit gates that is exponential in the number $n_q$ of qubits, thus
becoming rapidly unfeasible when increasing $n_q$. On the other hand,
pseudo-random states approximating to the desired accuracy the entanglement
properties of true random states may be generated efficiently, that is,
polynomially in $n_q$. In particular, quantum chaotic maps are efficient
generators of multipartite entanglement among the qubits, close to that
expected for random states.
This review discusses several aspects of 
the relationship between entanglement, randomness
and chaos. In particular, I will focus on the following items:
(i) the robustness of the entanglement generated by quantum chaotic maps 
when taking into account the unavoidable noise sources affecting a 
quantum computer; (ii) the detection of the entanglement of 
high-dimensional (mixtures of) random states, an issue also related to 
the question of the emergence of classicality in coarse grained 
quantum chaotic dynamics; (iii) the decoherence induced 
by the coupling of a system to a chaotic environment, that is, by 
the entanglement established between the system and the environment.
\end{abstract}

\section{Introduction}

Entanglement~\cite{bruss,pleniovirmani,horodecki},
arguably the most spectacular and counterintuitive
manifestation of quantum mechanics, is 
observed in composite quantum systems. It
signifies the existence of non-local correlations between
measurements performed on particles that have interacted in the past,
but now are located arbitrarily far away.
We say that a two-particle state $|\psi\rangle$
is entangled, or non-separable, if it cannot be written as a simple
tensor product 
$|k_1\rangle|k_2\rangle\equiv|k_1\rangle\otimes |k_2\rangle$ 
of two states
which describe the first and the second subsystems, respectively,
but only as a superposition of such states:
$|\psi\rangle=\sum_{k_1,k_2}c_{k_1k_2}|k_1\rangle|k_2\rangle$.
When two systems are entangled, it is not possible to assign them 
individual state vectors. 

The intriguing non-classical properties 
of entangled states were clearly illustrated by Einstein, Podolsky
and Rosen (EPR) in 1935~\cite{EPR}. These authors showed that quantum theory
leads to a contradiction, provided that we accept 
(i) the reality principle: If we can predict with certainty the value
of a physical quantity, then this value has physical reality, independently of
our observation;~\footnote{For example, if a system's
wave function $|\psi\rangle$ is an eigenstate of an operator 
$\hat{A}$, namely, $\hat{A} |\psi\rangle = a |\psi\rangle$,
then the value $a$ of the observable $A$ is, using the EPR language, 
an element of physical reality.}
(ii) the locality principle: If two systems are
causally disconnected, the result of any measurement performed on one
system cannot influence the result of a measurement performed on the
second system.~\footnote{Following the theory of relativity, we say 
that two measurement events are causally disconnected if
$(\Delta{x})^2>c^2(\Delta{t})^2$, where $\Delta{x}$ and $\Delta{t}$ are
the space and time separations of the two events in some inertial reference
frame and $c$ is the speed of light (the two events take place at
space-time coordinates $(x_1,t_1)$ and $(x_2,t_2)$, respectively, and
$\Delta{x}=x_2-x_1$, $\Delta{t}=t_2-t_1$).}
The EPR conclusion was that quantum mechanics is an
incomplete theory. 
The suggestion was that measurement is in reality a deterministic process, 
which merely appears probabilistic since some degrees of freedom 
(hidden variables) are not precisely known.
Of course, according to the standard interpretation 
of quantum mechanics there is no contradiction, since the
wave function is not seen as a physical object, but just as a mathematical
tool, useful to predict probabilities for the outcome of experiments.

The debate on the physical reality of quantum systems became
the subject of experimental investigation after the
formulation, in 1964, of Bell's inequalities~\cite{bell}. These
inequalities are obtained assuming the principles of realism and locality.
Since it is possible to devise situations in which quantum mechanics predicts a
violation of these inequalities, any experimental observation of such a
violation excludes the possibility of a local and realistic description of
natural phenomena. In short, Bell showed that the principles of realism and
locality lead to experimentally testable inequality relations in disagreement
with the predictions of quantum mechanics.

Many experiments have been performed in order to check Bell's inequalities; 
the most famous involved EPR pairs of photons and was performed by
Aspect and
coworkers in 1982~\cite{aspect}. 
This experiment displayed an unambiguous
violation of a Bell's inequality by tens of standard deviations and an
excellent agreement with quantum mechanics. More recently, other experiments
have come closer to the requirements of the ideal EPR scheme
and again impressive agreement with the predictions of quantum
mechanics has always been found. Nonetheless, there is no
general consensus as to whether or not these experiments
may be considered conclusive, owing to the limited
efficiency of detectors. If, for the sake of
argument, we assume  that the present results will not be
contradicted by future experiments with high-efficiency detectors, we must
conclude that Nature does not experimentally
support the EPR point of view. In summary, the
World is not locally realistic.

I should stress that there is more to learn from Bell's
inequalities and Aspect's experiments than merely a consistency test of
quantum mechanics. These profound results show us that \emph{entanglement is a
fundamentally new resource}, beyond the realm of classical physics,
and that it is possible to experimentally manipulate entangled states.
A major goal of quantum information science~\cite{qcbook,nielsen} 
is to exploit this resource 
to perform computation and communication tasks beyond classical capabilities.

Entanglement is central to many quantum communication protocols, 
including quantum dense coding~\cite{densecoding}, 
which permits transmission of two bits of classical
information through the manipulation of only one of two entangled qubits,
and quantum teleportation~\cite{teleportation}, which allows the transfer
of the state of one quantum system to another over an arbitrary distance.
Moreover, entanglement is a tool for secure communication~\cite{ekert}.
Finally, in the field of quantum computation entanglement allows
algorithms exponentially faster than any known classical 
computation~\cite{shor}.
For any quantum algorithm operating on pure states, the presence of
multipartite (many-qubit) entanglement is necessary to achieve an
exponential speedup over classical computation~\cite{jozsa}.
Therefore the ability to control high-dimensional 
entangled states is one of the basic requirements for constructing 
quantum computers.

Random numbers are important in classical computation, as probabilistic 
algorithms can be far more efficient than deterministic ones in solving
many problems~\cite{papadimitriou}.
Randomness may also be useful in quantum computation. 
\emph{Random pure states} of dimension $N$ 
are drawn from the uniform (Haar) measure on pure
states~\footnote{The Haar measure on the
unitary group $U(N)$ is the unique measure on pure $N$-level states
invariant under unitary transformations. It is a uniform,
unbiased measure on pure states. For a single qubit ($N=2$), it can 
be simply visualized as a uniform distribution on the 
\emph{Bloch sphere}~\cite{qcbook,nielsen}.
The generic state
of a qubit may be written as
\begin{equation}
|\psi\rangle=\cos\frac{\theta}{2}|0\rangle +
e^{i\phi}\sin\frac{\theta}{2}|1\rangle,
\quad
0\leq\theta\leq\pi, \ 0\leq\phi<2\pi,
\label{sphere}
\end{equation}
where the states of the \emph{computational basis}
$\{|0\rangle,|1\rangle\}$ are eigenstates of the Pauli operator
$\hat{\sigma}_z$. The qubit's state can be represented by a point on 
a sphere of unit radius, called the Bloch sphere. This sphere is 
parametrized by the angles $\theta,\phi$ and can be embedded in 
a three-dimensional space of Cartesian coordinates 
$(x=\cos\phi\sin\theta,y=\sin\phi\sin\theta,z=\cos\theta)$.\\
We also point out that ensembles of \emph{random mixed states}
are reviewed in~\cite{karol}.}
The entanglement content of random pure quantum states is almost
maximal~\cite{random-states,zyczkowski,hayden} and 
such states find applications in various quantum protocols, like
superdense coding of quantum states~\cite{harrow,hayden},
remote state preparation~\cite{bennett2005}, and the construction of
efficient data-hiding schemes~\cite{hayden2}.
Moreover, it has been argued that random evolutions
may be used to characterize the main aspects of noise sources affecting
a quantum processor~\cite{emerson1}.
Finally, random states may form the basis for a statistical theory
of entanglement. While it is very difficult to characterize 
the entanglement properties of a many-qubit state, a simplified
theory of entanglement might be possible for random states~\cite{hayden}.

The preparation of a random state or, equivalently, the implementation
of a random unitary operator mapping a fiducial $n_q$-qubit initial state,
say $|0\rangle\equiv |0\ldots 0\rangle
\equiv|0\rangle \otimes  
\ldots \otimes |0\rangle$, onto a typical (random) state,
requires a number of elementary
one- and two-qubit gates exponential in the number $n_q$ of qubits,
thus becoming rapidly unfeasible when increasing $n_q$.
On the other hand, pseudo-random states approximating to the desired
accuracy the entanglement properties of true random states may be
generated efficiently, that is, polynomially in
$n_q$~\cite{emerson1,emerson2,weinstein,plenio1,plenio2,znidaric}.
In a sense, pseudo-random states play in quantum 
information protocols a role analogous to pseudo-random numbers 
in classical information theory. 

Random states can be efficiently approximated by means of random one- and
two-qubit 
unitaries~\cite{emerson1,emerson2,weinstein,plenio1,plenio2,znidaric,gennaro}
or by deterministic dynamical systems (maps) in the regime of quantum 
chaos~\cite{haake,stoeckmann,saraceno,schack,georgeot,simone2002,caves,weinstein}.
These maps are known to exhibit certain statistical properties of 
random matrices~\cite{haake,stoeckmann} and are efficient generators 
of multipartite entanglement among the qubits, close to that expected 
for random states~\cite{caves,weinstein}.
Note that in this case deterministic instead of random one- and
two-qubit gates are implemented, the required randomness being 
provided by deterministic chaotic dynamics.
A related crucial question, which I shall discuss in this review, 
is whether the generated entanglement
is robust when taking into account unavoidable noise sources
affecting a quantum computer. That is, 
decoherence or imperfections in the quantum 
computer hardware~\cite{qcbook,varenna05}, 
that in general turn pure states into
mixtures, with a corresponding loss of quantum coherence
and entanglement content.

This paper reviews previous work concerning  
several aspects of the relationship between entanglement, randomness
and chaos. In particular, I will focus on the following items:
(i) the robustness of the entanglement generated by quantum chaotic maps
when taking into account the unavoidable noise sources affecting a
quantum computer (Sec.~\ref{sec:stabilitymultipartite}); 
(ii) the detection of the entanglement of
high-dimensional (mixtures of) random states, an issue also related to
the question of the emergence of classicality in coarse grained
quantum chaotic dynamics (Sec.~\ref{sec:detect}); 
(iii) the decoherence induced
by the coupling of a system to a chaotic environment, that is, by
the entanglement established between the system and the 
environment (Sec.~\ref{sec:chaoticenvironments}).
In order to make this paper accessible also to readers without a
background in quantum information science and/or in quantum chaos, 
basic concepts and tools concerning bipartite and multipartite entanglement, 
random and pseudo-random quantum states and quantum chaos maps 
are discussed in the remaining sections and appendixes.

\section{Bipartite entanglement}

\label{sec:bipartite}

\subsection{The von Neumann entropy}

In this section, we show that \emph{for pure states} $|\psi\rangle$ 
a good measure of bipartite entanglement exists: the von Neumann
entropy of the reduced density matrices. Given a state described 
by the density matrix $\rho$, its von Neumann entropy is defined as 
\begin{equation}
S(\rho)=-{\rm Tr}(\rho\log\rho). 
\end{equation}
Note that, here as in the rest of
the paper, all logarithms are base $2$ unless otherwise indicated.

First of all, a few definitions are needed:

\emph{Entanglement cost}: Let us assume that two communicating 
parties, Alice and Bob,
share many Einstein-Podolsky-Rosen (EPR) pairs~\footnote{A basis of
(maximally) entangled states for the two-qubit Hilbert space is provided 
by the four Bell states (EPR pairs)
\begin{equation}
|\phi^\pm\rangle=\frac{1}{\sqrt{2}}\,\big(|00\rangle\pm|11\rangle\big),
\quad
|\psi^\pm\rangle=\frac{1}{\sqrt{2}}\,\big(|01\rangle\pm|10\rangle\big).
\end{equation} 
}, 
say 
\begin{equation}
|\phi^+\rangle=\frac{1}{\sqrt{2}}\,\big(|00\rangle+|11\rangle\big),
\label{EPRstate}
\end{equation} 
and that they
wish to prepare a large number $n$ of copies of a given bipartite
pure state $|\psi\rangle$, using only local operations and classical
communication. If we call $k_{\min}$ the minimum number of EPR pairs
necessary to accomplish this task, we define the entanglement cost as the
limiting ratio $k_{\min}/n$, for $n\to\infty$.

\emph{Distillable entanglement}: Let us consider the reverse
process; that is, Alice and Bob share a large number $n$ of
copies of a pure state $|\psi\rangle$ and they wish to
concentrate entanglement, again using only local operations supplemented by
classical communication. If $k_{\max}'$ denotes the maximum number of EPR pairs
that can be obtained in this manner, we define the distillable
entanglement as the ratio $k_{\max}'/n$ in the limit $n\to\infty$.

It is clear that $k_{\max}' \leq k_{\min}$. Otherwise, we could
employ local operations and classical communication to create
entanglement, which is a non-local, purely quantum resource (it would be
sufficient to prepare $n$ states $|\psi\rangle$ from $k_{\min}$ EPR
pairs and then distill $k_{\max}'>k_{\min}$ EPR states). Furthermore, it is
possible to show that, asymptotically in $n$, the entanglement cost and the
distillable entanglement coincide and that the ratios $k_{\min}/n$ and
$k_{\max}'/n$ are given by the reduced single-qubit von Neumann entropies.
Indeed, we have
\begin{equation}
  \lim_{n\to\infty} \frac{k_{\min}}{n}
  =
  \lim_{n\to\infty} \frac{k_{\max}'}{n}
  = S(\rho_A) = S(\rho_B),
\end{equation}
where $S(\rho_A)$ and $S(\rho_B)$ are the von Neumann entropies of the reduced
density matrices
$\rho_A={\rm Tr}_{B}\big(|\psi\rangle\langle\psi|\big)$ and
$\rho_B={\rm Tr}_{A}\big(|\psi\rangle\langle\psi|\big)$,
respectively. Therefore, the process that changes $n$ copies of
$|\psi\rangle$ into $k$ copies of $|\phi^+\rangle$ is
asymptotically reversible. Moreover, it is possible to show that it is
faithful; namely, the change takes place with unit
fidelity when $n\to\infty$.\footnote{The fidelity $F$ provides a 
measure of the distance between two, generally mixed, quantum states
$\rho$ and $\sigma$:
\begin{equation}
  F(\rho,\sigma) = \left({\rm Tr} \sqrt{\rho^{1/2}\sigma\rho^{1/2}}\right)^2.
\end{equation}
The fidelity of a pure state $|\psi\rangle$
and an arbitrary state $\sigma$ is given by
\begin{equation}
F(|\psi\rangle,\sigma)={\langle \psi | \sigma | \psi \rangle},
\end{equation}
which is the square root of the overlap between $|\psi\rangle$ and $\sigma$.
Finally, the fidelity of two pure quantum states 
$|\psi_1\rangle$
and $|\psi_2\rangle$ is defined by
\begin{equation}
F(|\psi_1\rangle,|\psi_2\rangle)=
|\langle\psi_1|\psi_2\rangle|^2.
\end{equation}
We have $0\leq{}F\leq{}1$, with $F=1$
when $|\psi_1\rangle$ coincides with $|\psi_2\rangle$ and $F=0$ when
$|\psi_1\rangle$ and $|\psi_2\rangle$ are orthogonal.
For further discussions on this quantity see, 
e.g.,~\cite{qcbook,nielsen}. The average fidelity between 
random states is studied in~\cite{karolsommers}.} The proof of this result 
can be found in~\cite{bennett}. We can therefore
quantify the entanglement of a bipartite pure state 
$|\psi\rangle$ as
\begin{equation}
  E_{AB}(|\psi\rangle) = S(\rho_A) = S(\rho_B).
  \label{entbipure}
\end{equation}
It ranges from $0$ for a separable state to $1$ for maximally entangled
two-qubit states (the EPR states). Hence, it is common practice to say
that the entanglement of an EPR pair is $1$ \emph{ebit}. 

\subsection{The Schmidt decomposition}

\label{sec:schmidt}

The fact that $S(\rho_A)=S(\rho_B)$ is easily derived from the Schmidt
decomposition.

\emph{The Schmidt decomposition theorem}:
 Given a pure state
$|\psi\rangle\in\mathcal{H}=\mathcal{H}_A\otimes\mathcal{H}_B$ of a bipartite
quantum system, there exist orthonormal states $\{|i\rangle_A\}$ for 
$\mathcal{H}_A$ and $\{|i'\rangle_B\}$ for $\mathcal{H}_B$ such that
\begin{equation}
  |\psi\rangle =
  \sum_{i=1}^k
  \sqrt{p_i} \, |i\rangle_A |i'\rangle_B
  =
  \sqrt{p_1} \, |1\rangle_A |1'\rangle_B +
  \cdots +
  \sqrt{p_k} \, |k\rangle_A |k'\rangle_B ,
  \label{schmidtdec}
\end{equation}
with $p_i$ positive real numbers satisfying the condition $\sum_{i=1}^k p_i=1$
(for a proof of this theorem see, e.g.,~\cite{qcbook,nielsen}).

It is important to stress that the states $\{|i\rangle_A\}$ and
$\{|i'\rangle_B\}$ depend on the particular state $|\psi\rangle$ that we
wish to expand. 
The reduced density matrices $\rho_A=\sum_i p_i |i\rangle_A\,{}_A\langle i|$
and $\rho_B=\sum_i p_i |i'\rangle_B\,{}_B\langle i'|$ have the same 
non-zero eigenvalues. Their number is also the number $k$ of terms in the
Schmidt decomposition (\ref{schmidtdec}) and is known as the
\emph{Schmidt number}
(or the \emph{Schmidt rank})
of the state $|\psi\rangle$. A separable
pure state, which by definition can be written as
\begin{equation}
  |\psi\rangle = |\phi\rangle_A |\xi\rangle_B,
\end{equation}
has Schmidt number equal to one. Thus, we have the following entanglement
criterion: a bipartite pure state is entangled if and only if 
its Schmidt number is greater than one.
For instance, the Schmidt number of the EPR state (\ref{EPRstate}) 
is $2$.

It is clear from the Schmidt decomposition (\ref{schmidtdec}) that
\begin{equation}
S(\rho_A)=S(\rho_B)=-\sum_i p_i\log p_i. 
\end{equation}
If $N$, $N_A$ and $N_B$
denote the dimensions of the Hilbert spaces $\mathcal{H}$,
$\mathcal{H}_A$ and $\mathcal{H}_B$, with $N=N_AN_B$ and $N_A\le N_B$,
we have 
\begin{equation}
0\le E_{AB}(|\psi\rangle)=S(\rho_A)=S(\rho_B)\le \log N_A.
\end{equation}
A maximally entangled state of two subsystems has 
$N_A$ equally weighted terms in its Schmidt decomposition 
and therefore its entanglement content is
$\log{N_A}$ ebits.
For instance, the EPR state (\ref{EPRstate}) is a maximally 
entangled two-qubit state.
Note that a maximally entangled state $|\psi\rangle$ 
leads to a maximally mixed state $\rho_A$.

\subsection{The purity}

The purity of state described by the density matrix $\rho$ 
is defined as 
\begin{equation}
P(\rho)={\rm Tr}(\rho^2).
\end{equation}
We have 
\begin{equation}
P(\rho_A)=P(\rho_B)=\sum_{i}p_i^2.
\end{equation}
The purity is much easier to investigate analytically than the 
von Neumann entropy. Moreover, it provides the first non-trivial 
term in a Taylor series expansion of the von Neumann entropy about 
its maximum.\footnote{If we write $p_i=\frac{1+\epsilon_i}{N_A}$,
with $\epsilon_i\ll 1$ and $\sum_i \epsilon_i=0$, we obtain
\begin{equation}
S(\rho_A)\approx \log N_A -[N_A/(2\ln 2)]P(\rho_A).
\end{equation}}
The purity ranges from $1/N_A$ for maximally entangled states to
$1$ for separable states. 
One can also consider the \emph{participation ratio}
$\xi=\frac{1}{\sum_i p_i^2}$, which is the inverse of the purity. 
This quantity is bounded between $1$ and $N_A$ and is close to $1$
if a single term dominates the Schmidt decomposition 
(\ref{schmidtdec}), whereas $\xi=N_A$ if all terms in the decomposition 
have the same weight ($p_1=\cdots=p_{N_A}=1/N_A$).
The participation ratio $\xi$ represents the effective number of 
terms in the Schmidt decomposition.

\vspace{0.2cm}

A natural extension of the discussion of this section is to
consider \emph{bipartite mixed states},
$\rho=\sum_i p_i|\psi\rangle\langle\psi|$ $(\sum_i p_i=1$),
instead of pure states. However, mixed-state entanglement is
not as well understood as pure-state bipartite
entanglement and is the focus of ongoing research 
(for a review, see, e.g.,
Refs.~\cite{bruss,pleniovirmani,horodecki}).

By definition, a (generally mixed) state 
is said to be separable if it can be prepared by 
two parties (Alice and Bob) in a ``classical'' manner; that is, by
means of local operations and classical communication. This means that Alice and
Bob agree over the phone on the local preparation of the two subsystems $A$ and
$B$. Therefore, a mixed state is separable if and only if it can be written as
\begin{equation}
  \rho_{AB} = \sum_k p_k \, \rho_{Ak} \otimes \rho_{Bk}, \quad
  \hbox{with } p_k \ge 0 \hbox{ and }
  \sum_k p_k = 1,
  \label{sepdecomposition}
\end{equation}
where $\rho_{Ak}$ and $\rho_{Bk}$ are density matrices for the two subsystems. A
separable system always satisfies Bell's inequalities; that is, it
only contains classical correlations.
Given a density matrix $\rho_{AB}$, it is in general a non-trivial task
to prove whether a decomposition as in (\ref{sepdecomposition}) exists
or not~\cite{pleniovirmani,horodecki}. 
We therefore need separability criteria that are easier to test. 
Two useful tools for the detection of entanglement, the 
Peres criterion and entanglement witnesses, are reviewed in 
Appendix~\ref{app:separability}.

\section{Entanglement of random states}

\label{sec:entrandom}

A simple argument helps understanding why the bipartite entanglement
content of a pure random state $|\psi\rangle$ is almost maximal. 
In a given basis $\{|i\rangle\}$ the density matrix for the state 
$|\psi\rangle=\sum_i c_i |i\rangle$ is written as follows:
\begin{equation}
\rho_{ij} = \langle i | \psi\rangle
\langle \psi | j\rangle= c_i c_j^\star,
\end{equation}
where $c_i=\langle i | \psi \rangle$ are the components of the state
$|\psi\rangle$ in the $\{|i\rangle\}$ basis.
In the case of a random state the components are uniformly 
distributed, with amplitudes $c_i\approx 1 /\sqrt{N}$ and random phases.
Here $N$ is the Hilbert space dimension and the value
$1/\sqrt{N}$ of the amplitudes ensures that the wave vector 
$|\psi\rangle$ is normalized. The density matrix can therefore 
be written as 
\begin{equation}
\rho \approx 
\mathrm{diag}
\left(\frac{1}{{N}},\frac{1}{{N}},\ldots,
\frac{1}{{N}}\right) + 
\Omega,
\label{eq:rhototrandom}
\end{equation}
where  $\Omega$  is  a  ${N}\times{N}$  zero  diagonal
matrix  with  random complex  matrix  elements  of amplitude  $\approx
1/{N}$.
Suppose now that we partition the Hilbert space of the system into two
parts, $A$ and $B$, 
with dimensions  ${N}_A$  and  ${N}_B$,  where
${N}_A{N}_B={N}$. 
Without loss of generality, we take the first subsystem, $A$, to be 
the one with the not larger dimension: ${N}_A\le {N}_B$.
The reduced density matrix $\rho_A$ is defined as follows:
\begin{equation}
\rho_A=\mathrm{Tr}_B \rho=
\sum_{i_B} c_{i_A i_B} c^\star_{i_A^\prime i_B}
|i_A\rangle \langle i_A^\prime|,
\end{equation}
where  $|i\rangle =  |i_A i_B\rangle$. Using Eq.~(\ref{eq:rhototrandom}), 
we obtain
\begin{equation} 
\rho_A \approx \mathrm{diag}
\left(\frac{1}{{N}_A},\frac{1}{{N}_A},\ldots,
\frac{1}{{N}_A}\right) + 
\Omega_A,
\label{eq:rhoA}
\end{equation}
where $\Omega_A$ is a zero diagonal matrix with matrix elements of
$O\left(\sqrt{{N}_B}/{N}\ll
1/{N}_A\right)$ (sum   of
${N}_B\gg 1$ terms of order $1/{N}$ with random phases).
Neglecting $\Omega_A$ in (\ref{eq:rhoA}), the  
reduced von Neumann entropy of subsystem $A$ is given by  
$S(\rho_A)=\log ({N}_A)$, the maximum entropy 
that the subsystem $A$ can have.

\subsection{Page's formula}

The exact mean value 
$\langle E_{AB}(|\psi\rangle) \rangle$
of the bipartite entanglement  
is given by Page's formula, obtained~\cite{random-states}
by considering the ensemble of
random pure states drawn 
according to the Haar measure on $U(N)$: 
\begin{equation}
S_P\equiv \langle E_{AB}(|\psi\rangle) \rangle =
\langle S(\rho_A) \rangle=\langle S(\rho_B)\rangle=
\log{N}_A-\frac{{N}_A}{2{N}_B\ln 2},
\label{Epage}
\end{equation}
where $\langle\, \cdot\, \rangle$ denotes the (ensemble) average
over the uniform Haar measure.
For $\log{N}_A\gg 1$, $\langle E_{AB} \rangle$ 
is close to its maximum 
value $E_{AB}^{\hbox{max}}(|\psi\rangle)=\log{N}_A\gg 1$.
Note that, if we fix $N_A$ and let $N_B\to\infty$, then 
$\langle E_{AB}\rangle $ tends to $E_{AB}^{\hbox{max}}$.

Remarkably, if we consider the \emph{thermodynamic limit},
that is, we fix $N_A/N_B$ and let $N_A\to\infty$,
then the reduced von Neumann entropy concentrates around its 
average value (\ref{Epage}). This is a consequence of 
the so-called \emph{concentration of measure} phenomenon: 
the uniform measure on the $k$-sphere $\mathbb{S}^k$ in 
$\mathbb{R}^{k+1}$ (parametrized, for 
instance, by $k=N^2-2$ angles in the 
Hurwitz parametrization~\cite{pozniak,weinstein})
concentrates very strongly around the equator when $k$ is large:
Any polar cap smaller than a hemisphere has relative volume 
exponentially small in $k$. This observation implies, in particular,
the concentration of the entropy of the reduced density matrix
$\rho_A$ around its average value~\cite{sommers,hayden}.
This in turn implies that when the dimension $N$ of the quantum
system is large it is meaningful to apply statistical methods
and discuss typical 
(entanglement) behavior or random states, in the sense that
almost all random states behave in essentially the same way.

\subsection{Lubkin's formula}

For random states, the average value of the purity of the reduced
density matrices $\rho_A$ and $\rho_B$ is giben by
Lubkin's formula~\cite{lubkin}:
\begin{equation}
P_L\equiv \langle P(\rho_A)\rangle =
\langle P(\rho_B)\rangle =
\frac{N_A+N_B}{N_AN_B+1}.
\label{eq:lubkin}
\end{equation}
Nothe that, if we fix $N_A$ and let $N_B\to\infty$, then 
$P_L$ tends to its minimum value $1/N_A$. If we fix $N_A/N_B$ 
and let $N_A\to\infty$, then $P_L\to 0$.
For large $N$, the variance
\begin{equation}
\sigma_{P}^2=\langle P^2 \rangle - P_L^2 \approx \frac{2}{N^2},
\label{eq:variancelubkin}
\end{equation}
so that the relative standard deviation 
\begin{equation}
\frac{\sigma_{P}}{P_L} \approx \frac{\sqrt{2}}{N_A+N_B}
\end{equation}
tends to zero in the thermodynamic limit $N_A\to \infty$
(at fixed $N_A/N_B$). For \emph{balanced bipartition},
corresponding to $N_A=N_B=\sqrt{N}$, we have
\begin{equation}
\frac{\sigma_{P}}{P_L} =O\left(\frac{1}{\sqrt{N}}\right).
\end{equation}
Note that the fact that ${\sigma_{P}}/{P_L}\to 0$ when 
$N\to\infty$ is again a consequence of the concentration 
of measure phenomenon~\cite{sommers,hayden}.
A derivation of 
Eqs.~(\ref{eq:lubkin}) and (\ref{eq:variancelubkin})
is presented in Appendix~\ref{app:lubkin}.

\section{Pseudo-random states}

\label{sec:pseudo-random}

The generation of a random state $|\psi\rangle$ is exponentially hard.
Indeed, starting from a fiducial $n_q$-qubit state $|0\rangle=|0\ldots 0\rangle$
one needs to implement a typical (random) unitary operator $U$ (drawn from the 
Haar measure on $U({N}=2^{n_q})$) to obtain $|\psi\rangle=U|0\rangle$.
Since $U$ is determined by $4^{n_q}-2$ real parameters (for instance, the angles
of the Hurwitz parametrization~\cite{pozniak,weinstein}), 
its generation requires a sequence of elementary one- and two-qubit
gates whose length grows exponentially in the number of qubits.
Thus, the generation of random states is unphysical for a large number 
of qubits. 
On the other hand, one can consider the generation of pseudo-random states
that could reproduce the entanglement properties of truly random 
states~\cite{emerson1,emerson2,weinstein,plenio1,plenio2,znidaric}.
In Refs.~\cite{plenio1,plenio2} it has been proven that the average 
entanglement of a typical state can be reached to a fixed accuracy
within $O(n_q^3)$ elementary quantum gate. This proof holds for a
random circuit such that $U$ is the product, 
\begin{equation}
U=W_{t}W_{t-1}\cdots W_2W_1, 
\label{UCNOT}
\end{equation}
of a sequence of $t=O(n_q^3)$ 
two-qubit 
gates $W_k$ independently chosen at each step as follows: 
\begin{itemize}
\item
a pair of integers $(i,j)$, with $i\ne j$ 
is chosen uniformly at random from $\{1,...,n_q\}$;
\item
single-qubit gates (unitary transformations) 
$V_k^{(i)}$ and $V_k^{(j)}$, drawn independently 
from the Haar measure on $U(2)$, are applied; 
\item 
a ${\rm{CNOT}}^{(i,j)}$ gate with control qubit $i$ and 
target qubit $j$ is applied.\footnote{By definition, the 
${\rm{CNOT}}^{(i,j)}$
gate acts on the states 
$\{|xy\rangle\equiv |x\rangle_i\otimes|y\rangle_j=
|00\rangle,|01\rangle,|10\rangle,|11\rangle\}$ 
of the two-qubit computational basis as follows:
${\rm{CNOT}}^{(i,j)}$ turns
$|00\rangle$ into $|00\rangle$,
$|01\rangle$ into $|01\rangle$,
$|10\rangle$ into $|11\rangle$, and
$|11\rangle$ into $|10\rangle$.
The CNOT (controlled-NOT) gate flips the state of the
second (target) qubit if the first (control) qubit is in the state
$|1\rangle$ and does nothing if the first qubit is in the state
$|0\rangle$.
In short, ${\rm{CNOT}^{(i,j)}}|x\rangle_i |y\rangle_j=
|x\rangle_i |x\oplus y\rangle_j$, with $x,y\in\{0,1\}$ and $\oplus$
indicating addition modulo $2$.
By definition, given input bits $x,y$, the ${\rm XOR}$ gate outputs
$i\oplus j$. Therefore, the (quantum) ${\rm CNOT}$ gate acts on the states
of the computational basis as the (classical) ${\rm XOR}$ gate.
However, the ${\rm CNOT}$ gate, in contrast to the
${\rm XOR}$ gate, can also be applied to any superposition of the
computational basis states. The CNOT gate is the prototypical two-qubit
gate that is able to generate entanglement. For instance, CNOT maps
the separable state $|\psi\rangle=\frac{1}{\sqrt{2}}
(|0\rangle+|1\rangle)|0\rangle$ onto the maximally entangled (Bell) state
$|\phi^+\rangle={\rm CNOT}|\psi\rangle=
\frac{1}{\sqrt{2}}(|00\rangle+|11\rangle)$. 
Any unitary operation in the Hilbert space of $n_q$ qubits
can be decomposed into (elementary) one-qubit and two-qubit CNOT gates.}
\end{itemize}
Therefore, 
\begin{equation}
W_k={\rm{CNOT}}^{(i,j)}V_k^{(i)}V_k^{(j)}.
\label{WCNOT}
\end{equation}
The proof \cite{plenio1,plenio2}
that (\ref{UCNOT}) generates to within any desired accuracy the 
entanglement of a random state in a polynomial number of gates is based 
on the fact that the evolution of the purity of the two subsystems $A$
and $B$ can be described following a Markov chain approach, with 
gap in the Markov chain given by $\Delta(n_q)\geq  p(n_q)$, 
where $p(n_q)=O(n_q^{-2})$.
If the density matrix $\rho_t$ 
of the overall system is expanded in terms of Pauli 
matrices,
\begin{equation}
\rho_t=\sum_{\alpha_0,...,\alpha_{n_q-1}=0}^3
c_t^{(\alpha_0,...,\alpha_{n_q-1})} \sigma_0^{(\alpha_0)}\otimes...
\otimes \sigma_{n_q-1}^{(\alpha_{n_q-1})},
\end{equation}
where
\begin{equation}
c_t^{(\alpha_0,...,\alpha_{n_q-1})}= 
\frac{1}{N} {\rm Tr}\left(\sigma_0^{(\alpha_0)}\otimes...
\otimes \sigma_{n_q-1}^{(\alpha_{n_q-1})}\rho_t\right),
\end{equation}
with $\sigma_0\equiv I$, $\sigma_1\equiv\sigma_x$,
$\sigma_2\equiv\sigma_y$, $\sigma_3\equiv\sigma_z$,
we obtain
\begin{equation}
P(\rho_{A,t})=P(\rho_{B,t})=N_AN_B^2
\sum_{\alpha_0,...,\alpha_{n_A-1}=0}^3
[c_t^{(\alpha_0,...,\alpha_{n_A-1},0,...,0)}]^2.
\label{puritypauli}
\end{equation}
As usual in this paper, we consider $n_A+n_B=n_q$ and
$N_AN_B=N$.
In the case of model (\ref{WCNOT}), the evolution of the 
column vector
\begin{equation}
{\bf c_t^2}\equiv {}^t[(c_t^{(0,...,0)})^2,(c_t^{(0,...,0,1)})^2,...,
(c_t^{(3,...,3)})^2]
\end{equation}
is described \cite{plenio1,plenio2} by a Markov chain:
\begin{equation}
{\bf c_{t+1}^2}=M {\bf c_{t}^2}.
\end{equation}
Therefore, the asymptotic decay of purity is determined by the 
second largest eigenvalue $1-\Delta(n_q)$ 
of the matrix $M$ (the largest 
eigenvalue is the unit eigenvalue):
\begin{equation}
|\langle P(\rho_{A,t})\rangle -P_L|\asymp (1-\Delta(n_q))^t
=\exp\{[\ln(1-\Delta(n_q))]t\}.
\end{equation}
For model (\ref{WCNOT}), one obtains \cite{plenio1,plenio2}
$\Delta(n_q)\geq  p(n_q)$, with $p(n_q)=O(n_q^{-2})$.

Alternatively, a two-qubit gate different from 
${\rm{CNOT}}$~\cite{znidaric} 
or $W_t$ chosen from the $U(4)$ Haar measure and acting on a  
a pair $i,j$ of qubits ($i\ne j$)
randomly chosen at each step, can be 
used~\cite{znidaric,gennaro}.
Numerical results for these models~\cite{plenio1,plenio2,znidaric}
indicate that the above analytic bound is not optimal:
there is numerical evidence that $\Delta(n_q)=O(n_q^{-1})$ and  
therefore $O(n_q^2)$ steps are sufficient to generate the 
average entanglement of a random state.
A different strategy to efficiently approximate the  
entanglement content of random states is based on 
quantum chaotic maps~\cite{weinstein,rossini} and will be discussed in 
Sec.~\ref{sec:qcent}.

Finally, we note that the Markov chain approach allows an easy derivation
of Lubkin's formula (\ref{eq:lubkin}). Let us consider
$W_k$ chosen from the $U(4)$ Haar measure. 
In this case, the Markov matrix 
\begin{equation}
M=\frac{2}{n_q(n_q-1)}\sum_{i,j} M_{ij}^{(2)},
\end{equation}
with $M^{(2)}_{ij}$ acting non trivially
(differently from identity) only on the subspace spanned
by qubits $i$ and $j$ ($i,j=1,...,n_q$ and $i\ne j$). 
After averaging over the uniform Haar measure on
$U(4)$ one can see \cite{znidaric} that $M^{(2)}_{ij}$ preserves identity
($\sigma_i^{(0)}\otimes \sigma_j^{(0)}
\to \sigma_i^{(0)}\otimes \sigma_j^{(0)}$)
and uniformly mixes the other $15$ products $\sigma_i^{(\alpha_i)}\otimes
\sigma_j^{(\alpha_j)}$ \cite{znidaric}. 
Matrix elements are therefore 
$[M^{(2)}_{ij}]_{0,0}=1$,
$[M^{(2)}_{ij}]_{0,x}=[M^{(2)}_{ij}]_{x,0}=0$,
and $[M^{(2)}_{ij}]_{x,x'}=1/15$ for $x,x'\in \{1,...,15\}$.

The matrix $M$ has an
eigenvalue equal to $1$ (with multiplicity $2$) and all the other eigenvalues
smaller than $1$.
The eigenspace corresponding to the unit
eigenvalue of matrix $M$ is spanned by the column vectors
\begin{equation}
v_0={}^t(1,0,...,0),\;\;
v_1={}^t(0,1,...,1). 
\end{equation}
The asymptotic equilibrium
state $c^2(\infty)=\lim_{t\to \infty}c^2(t)$
is however uniquely
determined by the constraints ${\rm Tr}(\rho_t)=1$ and
${\rm Tr}(\rho_t^2)=1$, which impose
\begin{equation}
x_0\equiv [c_t^{(0,...,0)}]^2=\frac{1}{N^2},\;\;
x_1\equiv \sum_{\alpha_0,...,\alpha_{n_q-1}\ne (0,...,0)}
[c_t^{(\alpha_0,...,\alpha_{n_q-1})}]^2=\frac{N-1}{N^2}.
\end{equation}
Finally, we obtain
\begin{equation} 
c^2(\infty)=x_0 v_0+x_1\frac{1}{N^2-1}v_1=
\frac{1}{N^2}\,{}^{t}(1,\frac{N-1}{N^2-1},...,
\frac{N-1}{N^2-1})
\end{equation}
and, after substitution of the components of this vector into 
Eq.~(\ref{puritypauli}),
\begin{equation}
P(\rho_{A,t})=\frac{N_AN_B^2}{N^2}\left[1+
(N_A^2-1)\frac{N-1}{N^2-1}\right],
\end{equation}
which immediately leads to Lubkin's formula (\ref{eq:lubkin}).

\section{Multipartite entanglement}

\label{sec:multipartite}

The characterization and quantification of multipartite entanglement is a
challenging open problem in quantum information science and many different
measures have been proposed~\cite{pleniovirmani,horodecki}.
To grasp the difficulty of the problem, let us suppose to have $n$ 
parties composing the system we wish to analyze. In order to obtain a 
complete characterization of multipartite entanglement, we should take 
into account all possible non-local correlations among all parties. 
It is therefore clear that the number of measures needed to 
fully quantify multipartite entanglement
grows exponentially with the number of qubits.
Therefore, in Ref.~\cite{facchi} it has been proposed to 
characterize multipartite
entanglement by means of a function rather than with a single measure.
The idea is to look at the probability density function of bipartite
entanglement between all possible bipartitions of the system.
For pure states the bipartite entanglement is the von Neumann entropy
of the reduced density matrix of one of the two subsystems:
$E_{AB}(|\psi\rangle)=S(\rho_A)=S(\rho_B)$.

It is instructive to consider the smallest non-trivial instance 
where multipartite entanglement can arise: the three-qubit case. 
Here we have three possible bipartitions, with $n_A=1$ qubit and 
$n_B=2$ qubits. For a GHZ state~\cite{GHZstate},
\begin{equation} 
|{\rm GHZ}\rangle=\frac{1}{\sqrt{2}}(|000\rangle+|111\rangle),
\end{equation}
we obtain $\rho_A=\frac{I}{2}$ for all bipartitions, and therefore 
\begin{equation}
p(E_{AB})=\delta_{E_{AB},1},
\end{equation}
namely there is maximum multipartite entanglement, fully
distributed among the three qubits.~\footnote{The problem of 
finding, for a generic number  $n_q$ of 
qubits, maximally multipartite entangled states, that is, 
pure states for which the entanglement is maximal for each
bipartition, is discussed in~\cite{parisi}.}  Note that in this case
$\rho_B=\frac{1}{2}(|00\rangle\langle 00|+ 
|11\rangle\langle 11|)$ is separable and therefore the 
pairwise entanglement between any two qubits is equal to zero.

For a W state~\cite{Wstate},
\begin{equation} 
|{\rm W}\rangle=\frac{1}{\sqrt{2}}(|100\rangle+
|010\rangle+|001\rangle),
\end{equation}
we obtain $\rho_A=\frac{2}{3}|0\rangle\langle 0|+
\frac{1}{3}|1\rangle\langle 1|$ for all bipartitions, and therefore 
\begin{equation}
p(E_{AB})=\delta_{E_{AB},\bar{E}},
\end{equation}
where $\bar{E}=-\frac{2}{3}\log\left(\frac{2}{3}\right)
-\frac{1}{3}\log\left(\frac{1}{3}\right)\approx 0.92$,
namely the distribution $p(E_{AB})$ is peaked but the amount of 
multipartite entanglement is not maximal.

As a last three-qubit example, let us consider 
the state 
\begin{equation} 
|\psi\rangle=\frac{1}{\sqrt{2}}(|000\rangle+
|110\rangle)=\frac{1}{\sqrt{2}}(|00\rangle+|11\rangle)
|0\rangle,
\end{equation}
where the first two qubits are in a maximally entangled (Bell) state, 
while the third one is factorized. 
In this case, $\rho_A=\frac{I}{2}$ if subsystem $A$ is one of the 
first two qubits, $\rho_A=|0\rangle\langle 0|$ otherwise. Hence,
\begin{equation}
p(E_{AB})= \frac{2}{3}\delta_{E_{AB},1}+\frac{1}{3}\delta_{E_{AB},0},
\end{equation}
namely the entanglement can be large but the variance of the distribution 
$p(E_{AB})$ is also large.

For sufficiently large systems (${N} = 2^{n_q} \gg 1$),
it is reasonable to consider only balanced bipartitions,
i.e., with $n_A = n_B$ ($n_A+n_B=n_q$),
since the statistical weight of unbalanced ones ($n_A\ll n_B$)
becomes negligible~\cite{facchi}. 
If the probability density has a large mean value
$\langle E_{AB} \rangle \sim n_q$ ($\langle \, \cdot \, \rangle$ denotes
the average over balanced bipartitions) and small relative standard
deviation $\sigma_{AB}/\langle E_{AB}\rangle \ll 1$, we can conclude
that genuine multipartite entanglement is almost maximal
(note that $E_{AB}$ is bounded within the interval $[0,n_q]$).
As I shall discuss in Sec.~\ref{sec:qcent}, 
this is the case for random states~\cite{facchi} 
(see also Ref.~\cite{kendon}).\footnote{For 
a $n_q$-qubit GHZ state, $|{\rm GHZ}\rangle=\frac{1}{\sqrt{2}}
(|0...0\rangle+|1...1\rangle)$, the distribution $p(E_{AB})$ is 
peaked but at a small and essentially $n_q$-independent value, 
while for the cluster states~\cite{briegel}
the average entanglement $\langle E_{AB} \rangle$
is large and increases with $n_q$ but also the variance 
is large~\cite{facchi}.}

\section{Quantum chaos map and entanglement}

\label{sec:qcent}

The relation between chaos and entanglement is discussed, 
for instance, in~\cite{furuya,miller,lakshminarayan1,
lakshminarayan2,lakshminarayan3}.
More specifically, the use of quantum chaos for efficient
and robust generation of pseudo-random states carrying large
multipartite entanglement is nicely illustrated by the example
of the quantum sawtooth map~\cite{benenti01,varenna05,qcbook}.
This map is described by
the unitary Floquet operator $\hat{U}$:
\begin{equation}
|\psi_{t+1}\rangle = \hat{U}|\psi_t\rangle =
e^{-iT\hat{n}^{2}/2} \, e^{ik(\hat{\theta} -\pi)^{2}/2} 
|\psi_t\rangle ,
\label{eq:quantmap}
\end{equation}
where $\hat{n} = -i \, \partial/\partial \theta$,
$[\hat{\theta},\hat{n}]=i$ (we set $\hbar=1$) and the discrete
time $t$ measures the number of map iterations.
In the following I will always consider map (\ref{eq:quantmap})
on the torus $0 \leq \theta < 2 \pi$, $- \pi \leq p < \pi$,
where $p = T n$.
With an $n_q$-qubit quantum computer we are able to simulate the
quantum sawtooth map with ${N} = 2^{n_q}$ levels;
as a consequence, $\theta$ takes $N$ equidistant
values in the interval $0 \leq \theta < 2 \pi$, while
$n$ ranges from $-{N}/2$ to ${N}/2 -1$
(thus setting $T=2\pi/{N}$).
We are in the quantum chaos regime for map (\ref{eq:quantmap})
when $K\equiv kT >0$ or $K<-4$; in particular, in the following
I will focus on the case $K=1.5$.

There exists an efficient quantum algorithm for simulating
the quantum sawtooth map~\cite{benenti01,qcbook}.
The crucial observation is that the operator $\hat{U}$ in
Eq.~\ref{eq:quantmap} can be written as the product of two operators:
$\hat{U}_{k}= e^{ik(\hat{\theta}-\pi)^{2}/2}$
and $\hat{U}_{T}=e^{-iT\hat{n}^{2}/2}$,
that are diagonal in the $\theta$- and in the $n$-representation,
respectively. Therefore, the most convenient way to classically simulate
the map is based on the forward-backward fast Fourier
transform between $\theta$ and $n$ representations, and requires
$O({N}\log {N})$ operations per map iteration.
On the other hand, quantum computation exploits its capacity
of vastly parallelize the Fourier transform, thus requiring only
$O((\log {N})^2)$ one- and two-qubit gates
to accomplish the same task~\cite{benenti01,qcbook}.
In brief, the resources required by the quantum
computer to simulate the sawtooth map are
only logarithmic in the system size ${N}$, thus admitting an
exponential speedup, as compared to any known classical computation.
The sawtooth map and the quantum algorithm for its simulation
are discussed in details in Appendix~\ref{sec:sawmap}.

\begin{figure}
  \begin{center}
    \includegraphics[width=10.0cm]{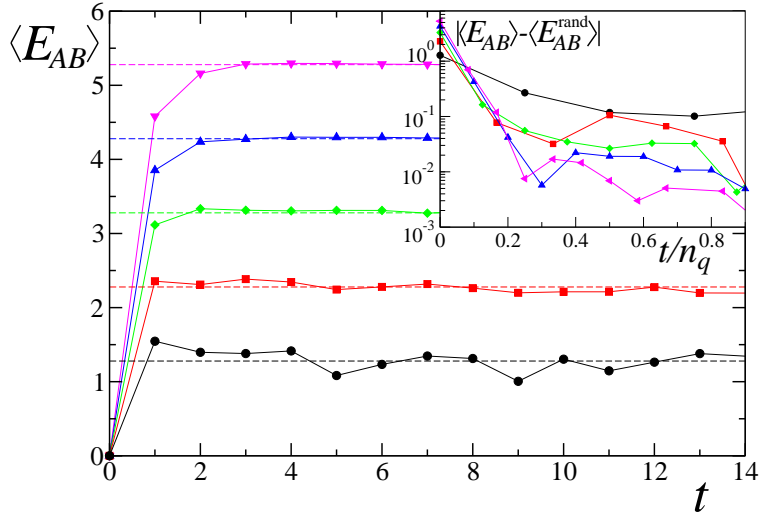}
    \caption{Time evolution of the average
      bipartite entanglement of a quantum state,
      starting from a state of the computational basis
      (eigenstate of the momentum operator $\hat{n}$),
      and recursively applying the quantum sawtooth map~(\ref{eq:quantmap})
      at $K=1.5$ and, from bottom to top, $n_q= 4, 6, 8, 10, 12$.
      Dashed lines show the theoretical values of
      Eq.~(\ref{eq:entpure}).
      Inset: convergence of $\langle E_{AB} \rangle (t)$
      to the asymptotic value
      $\langle E^{\, {\rm rand}}_{AB} \rangle$
      in Eq.~(\ref{eq:entpure}); time axis is rescaled
      with $1/n_q$. This figure is taken from Ref.~\cite{rossini}.}
    \label{fig:EntGen}
  \end{center}
\end{figure}

Let us first compute the average bipartite entanglement
$\langle E_{AB} \rangle$ as a function
of the number $t$ of iterations of map~(\ref{eq:quantmap}).
Numerical data in Fig.~\ref{fig:EntGen} exhibit a fast convergence,
within a few kicks, of this quantity to the value
\begin{equation}
\langle E^{\, {\rm rand}}_{AB} \rangle=
\frac{n_q}{2} - \frac{1}{2 \ln 2}
\label{eq:entpure}
\end{equation}
expected for a random state according to Page's 
formula~\cite{random-states} (note that 
this result is obtained from Eq.~(\ref{Epage}) in the special 
case $n_A=n_B$).
Precisely, as shown in the inset of Fig.~\ref{fig:EntGen},
$\langle E_{AB} \rangle$ converges exponentially fast to
$\langle E^{\,{\rm rand}}_{AB} \rangle$, with the time scale
for convergence $\propto n_q$.
Therefore, the average entanglement
content of a true random state is reached to a fixed accuracy
within $O(n_q)$ map iterations, namely $O(n_q^3)$
quantum gates.
I stress that in our case a deterministic map,
instead of random one- and two-qubit
gates as in Ref.~\cite{plenio1,plenio2,znidaric,gennaro},
is implemented. Of course, since the overall Hilbert space
is finite, the above exponential decay in a deterministic map is possible
only up to a finite time and the maximal accuracy drops exponentially
with the number of qubits.
I also note that, due to the quantum chaos regime,
properties of the generated pseudo-random state do not depend
on initial conditions, whose characteristics may even be very far
from randomness (e.g., simulations of Fig.~\ref{fig:EntGen},
start from a completely disentangled state).

\begin{figure}
  \begin{center}
    \includegraphics[width=10.0cm]{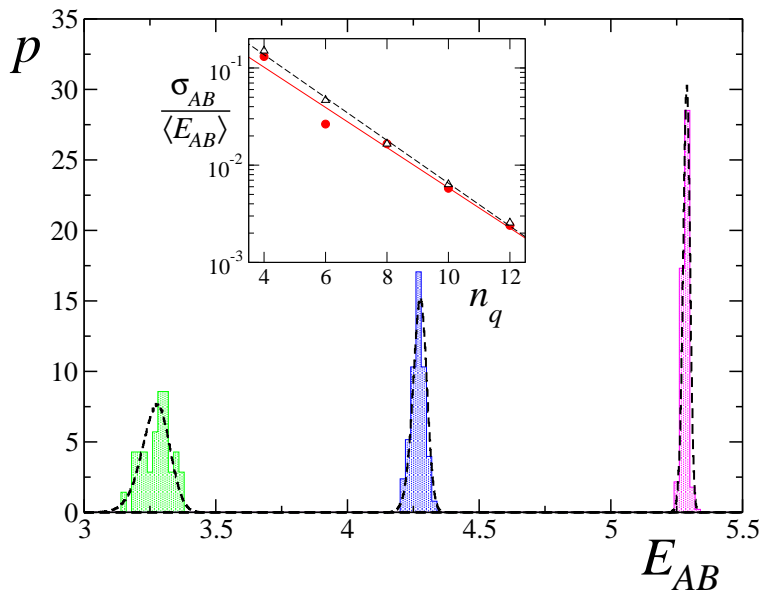}
    \caption{Probability density
      function of the bipartite von Neumann
      entropy over all balanced bipartitions for the state 
      $|\psi_t\rangle$,
      after $30$ iterations of map~(\ref{eq:quantmap}) at $K=1.5$.
      Various histograms are for different numbers of qubits:
      from left to right $n_q = 8, 10, 12$;
      dashed curves show the corresponding probabilities for random states.
      Inset: relative standard deviation
      $\sigma_{AB} / \langle E_{AB} \rangle$ as a function of $n_q$
      (full circles) and best exponential fit
      $\sigma_{AB} / \langle E_{AB} \rangle \sim e^{- 0.48 \, n_q}$
      (continuous line); data and best exponential fit
      $\sigma_{AB} / \langle E_{AB} \rangle \sim e^{- n_q / 2}$
      for random states are also shown (empty triangles, dashed line).
      This figure is taken from Ref.~\cite{rossini}.}
    \label{fig:Isto_eps0}
  \end{center}
\end{figure}

As discussed above, multipartite entanglement should
generally be described in terms of a function, rather than by
a single number. I therefore show in
Fig~\ref{fig:Isto_eps0} the probability density function
$p(E_{AB})$ for the entanglement of all possible balanced
bipartitions of the state $|\psi_{t=30}\rangle$.
This function is sharply peaked around
$\langle E^{\,{\rm rand}}_{AB}\rangle$,
with a relative standard deviation $\sigma_{AB} / \left< E_{AB} \right>$
that drops exponentially with $n_q$ (see
the inset of Fig.~\ref{fig:Isto_eps0})
and is small ($\sim 0.1$) already at $n_q=4$.
For this reason, we can conclude that multipartite entanglement is
large and that it is reasonable to use the first moment
$\langle E_{AB} \rangle$ of $p(E_{AB})$ for its characterization.
The corresponding probability
densities for random states is also calculated 
(dashed curves in Fig.~\ref{fig:Isto_eps0});
their average values and variances are in agreement with the values
obtained from states generated by the sawtooth map.
As we have remarked in Sec.~\ref{sec:entrandom}, 
the fact that for random states the distribution $p(E_{AB})$ is
peaked around a mean value close to the maximum achievable value
$E_{AB}^{\rm max}=n_q/2$ is a manifestation of the concentration
of measure phenomenon in a multi-dimensional Hilbert 
space~\cite{sommers,hayden}.

\section{Stability of multipartite entanglement}

\label{sec:stabilitymultipartite}

In order to assess the physical significance of the
generated multipartite entanglement, it is crucial to
study its stability when realistic noise is taken into account.
Hereafter I model quantum noise by means of unitary noisy gates,
that result from an imperfect control of the quantum computer
hardware~\cite{cirac95}. The noise model of 
Ref.~\cite{rossini04} is followed.
One-qubit gates can be seen as rotations of the Bloch sphere
about some fixed axis; I assume that unitary errors slightly tilt
the direction of this axis by a random amount.
Two-qubit controlled phase-shift gates are diagonal in the
computational basis; I consider unitary perturbations by
adding random small extra phases on all the
computational basis states.
Hereafter I assume that each noise parameter $\varepsilon_i$ is randomly
and uniformly distributed in the interval 
$[-\varepsilon, +\varepsilon]$;
errors affecting different quantum gates are also supposed to be
completely uncorrelated: every time we apply a noisy gate, noise
parameters randomly fluctuate in the (fixed) interval 
$[-\varepsilon, +\varepsilon]$.

Starting from a given initial state $|\psi_0\rangle$,
the quantum algorithm for simulating the sawtooth
map in presence of unitary noise gives an output state
$|\psi_{\varepsilon_I,t}\rangle$ that differs from
the ideal output $|\psi_t\rangle$. Here
$\varepsilon_I=(\varepsilon_1,\varepsilon_2,...,\varepsilon_{n_d})$ stands for
all the $n_d$ noise parameters
$\varepsilon_i$, that vary upon the specific noise configuration
($n_d$ is proportional to the number of gates).
Since we do not have any a priori knowledge of the
particular values taken by the parameters $\varepsilon_i$,
the expectation value of any observable $A$ for our
$n_q$-qubit system will be given by
${\rm Tr} [\rho_{\varepsilon,t} A]$, where the density matrix
$\rho_{\varepsilon,t}$ is obtained after averaging over noise:
\begin{equation}
\rho_{\varepsilon,t} =
\left(\frac{1}{2\varepsilon}\right)^{n_d}
\int d \varepsilon_I
|\psi_{\varepsilon_I,t}\rangle 
\langle\psi_{\varepsilon_I,t}| \, .
\label{eq:rhomatr}
\end{equation}
The integration over $\varepsilon_I$ is estimated numerically
by summing over $\mathcal{N}$ random realizations of noise,
with a statistical error vanishing in the limit $\mathcal{N}\to \infty$.
The mixed state $\rho_{\varepsilon}$ may also arise as a consequence
of non-unitary noise; in this case Eq.~(\ref{eq:rhomatr})
can also be seen as an unraveling of 
$\rho_{\varepsilon}$ into stochastically
evolving pure states 
$|\psi_{\varepsilon_I}\rangle$, each evolution
being known as a quantum trajectory~\cite{plenioknight,brun,carlo1,carlo2}.

I now focus on the entanglement content of 
$\rho_{\varepsilon,t}$.
Unfortunately, for a generic mixed state of $n_q$ qubits,
a quantitative characterization of entanglement
is not known, neither unambiguous~\cite{pleniovirmani,horodecki}.
Anyway, it is possible to give numerically accessible
lower and upper bounds
for the bipartite {\it distillable entanglement}
$E_{AB}^{(D)} (\rho_{\varepsilon})$:
\begin{equation}
\max \left\{ S(\rho_{\varepsilon,A}) - 
S(\rho_\varepsilon), 0 \right\} \leq
E_{AB}^{(D)} (\rho_{\varepsilon}) 
\leq \log \| \rho_{\varepsilon}^{T_B} \| \, ,
\label{eq:entbounds}
\end{equation}
where $\rho_{\varepsilon,A} = 
{\rm Tr}_B (\rho_\varepsilon)$ and
$\| \rho_{\varepsilon}^{T_B} \| \equiv {\rm Tr}
\sqrt{(\rho_\varepsilon^{T_B})^\dagger \, 
\rho_{\varepsilon}^{T_B}}$
denotes the trace norm of the partial transpose
of $\rho_{\varepsilon}$ with respect to party $B$
(see Appendix~\ref{app:peres} for the definition of the 
partial transposition operation).

\begin{figure}
  \begin{center}
    \includegraphics[width=12.0cm]{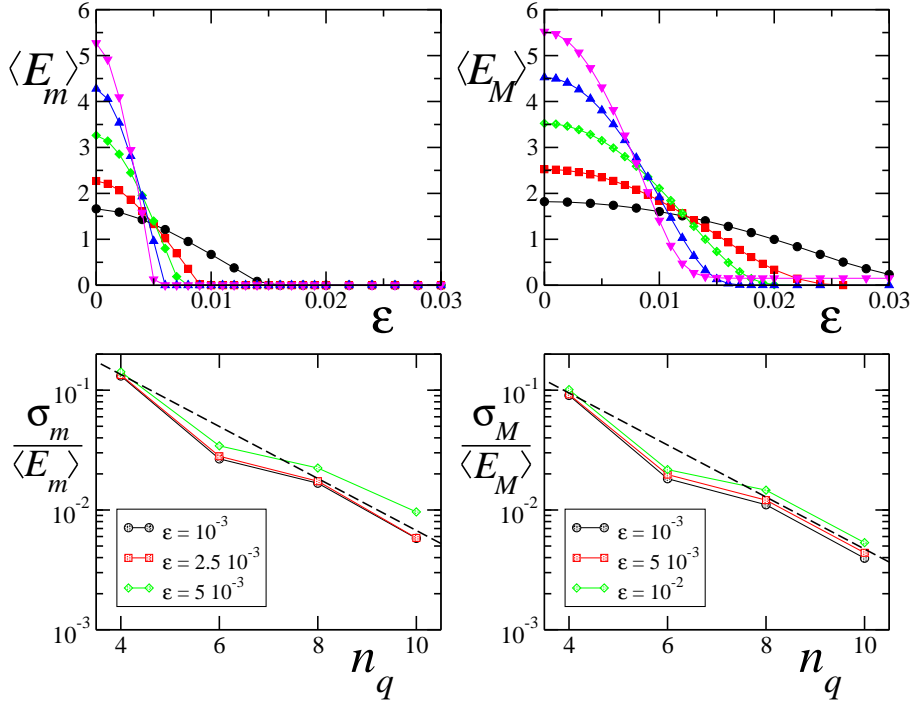}
    \caption{Upper graphs: lower $\langle E_m \rangle$ (left panel)
      and upper bound $\langle E_M \rangle$
      (right panel) for the bipartite distillable entanglement
      as a function of the noise strength at time $t=30$.
      Various curves stand for different numbers of qubits:
      $n_q=$ 4 (circles), 6 (squares), 8 (diamonds),
      10 (triangles up), and 12 (triangles down).
      Lower graphs: relative standard deviation of the probability density
      function for distillable entanglement over all balanced bipartitions
      as a function of $n_q$, for different noise strengths 
      $\varepsilon$.
      Dashed lines show a behavior
      $\sigma / \left< E \right> \sim e^{-n_q/2}$ and are
      plotted as guidelines. This figure is taken from Ref.~\cite{rossini}.}
    \label{fig:Ent_dest_nqvar}
  \end{center}
\end{figure}

In practice,  the quantum algorithm for the
quantum sawtooth map is simulated in the chaotic regime
with noisy gates and
the two bounds in Eq.~(\ref{eq:entbounds}) for the bipartite
distillable
entanglement of the mixed state $\rho_{\varepsilon,t}$,
obtained after averaging over $\mathcal{N}$ noise realizations,
are evaluated.
A satisfactory convergence for the lower and the upper bound
is obtained after $\mathcal{N}\sim \sqrt{N}$ and
$\mathcal{N}\sim N$ noise realizations, respectively.
The first moment of the lower ($E_m$) and the upper ($E_M$)
bound for the bipartite distillable entanglement
is shown as a function of the imperfection strength
in Fig.~\ref{fig:Ent_dest_nqvar}, upper panels.
The various curves are for different numbers $n_q$ of qubits;
$\mathcal{N}$ depends on $n_q$ and is large enough
to obtain negligible statistical errors (smaller than the size
of the symbols).
The relative standard deviation of the probability density
function (over all balanced bipartitions) for the 
bipartite distillable entanglement is shown in 
the lower panels of Fig.~\ref{fig:Ent_dest_nqvar}.
Like for pure states, we notice an exponential drop with $n_q$;
the distribution width slightly broadens when increasing
imperfection strength $\varepsilon$.
We can therefore conclude that an average value of the bipartite
bipartite 
distillable entanglement close to the ideal case $\varepsilon=0$
implies that multipartite entanglement is stable.

\begin{figure}
  \begin{center}
    \includegraphics[width=11.0cm]{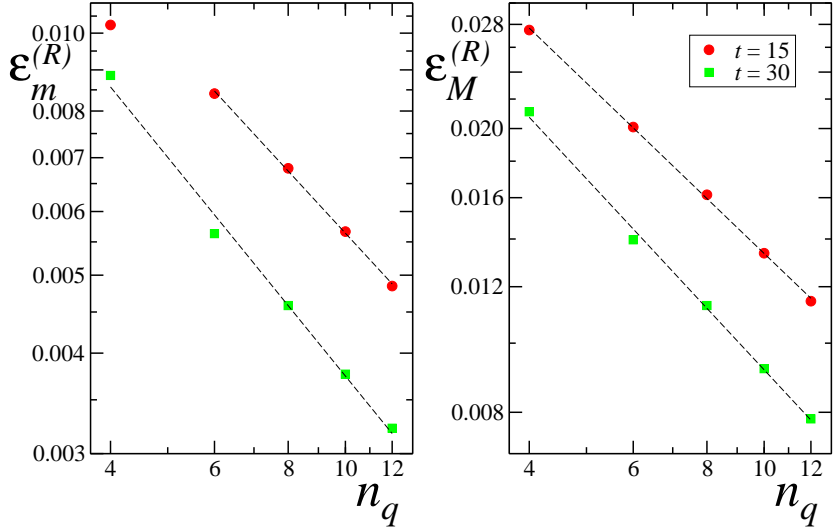}
    \caption{Perturbation strength at which the bounds of multipartite
      entanglement halve (lower bound on the left panel, upper bound
      on the right panel), as a function of the number of qubits.
      Dashed lines are best power-law fits of
      numerical data: 
      $\varepsilon^{(R)} \sim n_q^{-0.79 \pm 0.01}$ at $t=15$,
      $\varepsilon^{(R)} \sim n_q^{-0.9 \pm 0.01}$ at $t=30$, for both
      lower and upper bounds. This figure is taken from Ref.~\cite{rossini}.}
    \label{fig:EScal_eps_nq}
  \end{center}
\end{figure}

In order to quantify the robustness of multipartite entanglement
with the system size, let us define a perturbation strength threshold
$\varepsilon^{(R)}$ at which the distillable entanglement bounds drop
by a given fraction, for instance to $1/2$, of their 
$\varepsilon=0$ value,
and analyze the behavior of $\varepsilon^{(R)}$ as a function of
the number of qubits.
Numerical results are plotted in Fig.~\ref{fig:EScal_eps_nq};
both for lower and upper bounds we obtain a power-law scaling
close to
\begin{equation}
\varepsilon^{(R)} \sim 1/n_q \, .
\label{eq:epsscaling}
\end{equation}

It is possible to give a semi-analytical proof of the
scaling~(\ref{eq:epsscaling}) for the lower bound measure,
that is based on the quantum Fano inequality~\cite{schumacher96},
which relates the entropy $S(\rho_{\varepsilon})$ to the fidelity
$F = \langle \psi_t \vert \rho_{\varepsilon,t}
\vert \psi_t \rangle$:
\begin{equation}
S(\rho_{\varepsilon})\,\lesssim\, h(F) + (1 - F) \, \log (N^2 - 1),
\label{eq:qfano}
\end{equation}
where $h(x) = -x \log (x) - (1-x) \log (1-x)$ is the binary
Shannon entropy.
Since 
$F \simeq e^{- \gamma \varepsilon^2 n_g t}$~\cite{rossini04,bettelli04},
with $\gamma \sim 0.28$ and $n_g = 3 n_q^2 + n_q$ being the number of
gates required for each map step, we obtain, for 
$\varepsilon^{2} n_g t \ll 1$,
\begin{equation}
S (\rho_{\varepsilon}) \, \le \, \gamma \varepsilon^2 n_g t \left[
 - \log ( \gamma \varepsilon^2 n_g t) + 2 n_q + \frac{1}{\ln 2} \right].
\label{eq:entrofano}
\end{equation}
For sufficiently large systems the second term dominates (for $n_q =12$
qubits, $t=30$ and $\varepsilon \sim 5 \times 10^{-3}$ the other terms
are suppressed by a factor $\sim 1/ 10$) and, to a first
approximation, we can only retain it.
On the other hand, an estimate of the reduced entropy $S(\rho_A)$
is given by the bipartite entropy~(\ref{eq:entpure})
of a pure random state~\cite{random-states}.
Therefore, from Eq.~(\ref{eq:entbounds}) we obtain the following
expression for the lower bound of the distillable entanglement:
\begin{equation}
E_{AB}^{(D)} (\rho_{\varepsilon}) \, \ge \,
\frac{n_q}{2} - \frac{1}{2 \ln 2} - 
6 \gamma n_q^3 \varepsilon^2 t \,.
\label{eq:fanoscaling}
\end{equation}
From the threshold definition
$E_{AB}^{(D)} 
(\rho_{\varepsilon^{(R)}}) = \frac{1}{2} E_{AB}^{(D)} (\rho_{0})
=\frac{1}{2}S(\rho_A)$
we get the scaling~(\ref{eq:epsscaling}), that is valid
when $n_q \gg 1$:
\begin{equation}
\varepsilon^{(R)}_m \sim 1/\sqrt{24 \, \gamma \, n_q^2 \, t}.
\end{equation}
Notice that, for small systems as the ones that can be numerically
simulated (see data in Fig.~\ref{fig:EScal_eps_nq}),
the first term of Eq.~(\ref{eq:entrofano}) may introduce remarkable
logarithmic deviations from the asymptotic power-law behavior.
At any rate, the scaling derived from Eq.~(\ref{eq:fanoscaling}) is
in good agreement with the above shown numerical data, and also reproduces
the prefactor in front of the power-law decay
(\ref{eq:epsscaling}) up to a factor of two.

\section{Detecting entanglement of random states}

\label{sec:detect}

The entanglement content of high-dimensional random pure states is almost
maximal: nevertheless, in this section I will demonstrate that, due 
to the complexity of such states,
the detection of their entanglement is rather difficult~\cite{slovenia}.

\subsection{Random states and the quantum to classical transition}

A motivation to the study of the detection of random states
comes from considerations of the quantum to classical transition.
Since random states carry a lot of entanglement and
entanglement has no analogue in classical mechanics,
one can immediately conclude that random states are highly non-classical.
On the other hand, as we have seen in Sec.~\ref{sec:qcent},
pseudo-random states with properties close
to those of true random states can be efficiently generated by
dynamical systems (maps) in the regime of quantum chaos.
In such chaotic maps the classical limit is recovered when
the number of levels $N\to\infty$. 
Therefore, one can argue that for large random states,
i.e., in the limit $N\to \infty$, the quantum expectation value
of an operator with a well defined classical limit will be close to 
its classical microcanonical average. According to this picture
random states in a way ``mimic'' classical microcanonical density.
Expectation values are therefore close to the classical ones.

At first sight this is in striking contrast with the almost maximal
entanglement of such states. However, as I shall discuss in the
following, the contradiction is only apparent~\cite{slovenia}.
The detection of entanglement for a random state appears very difficult
at large $N$, as it would demand the control of very 
finely interwoven degrees of freedom and a measurement
resolution inversely proportional to $N$,
which seems hardly feasible experimentally.
Therefore, as far as the detection of
entanglement is concerned, high dimensional random states are
effectively classical.~\footnote{Of course, this remark does
not call into question the utility of high dimensional 
random states for quantum information processing.}

Moreover, coarse graining naturally appears.
For instance, one could repeat several times the measurement of
an entanglement witness (the definition of entanglement witness
is provided in Appendix~\ref{sec:witness})
for a random state and the prepared
random state would be different from time to time due to
unavoidable experimental imperfections. Let us model this problem
by considering mixtures of $m$ pure random states, namely
\begin{equation}
\rho=\sum_{i=1}^m \frac{1}{m} \ket{\psi_i}\bra{\psi_i},
\label{eq:rho}
\end{equation}
where the $\ket{\psi_i}$ are mutually independent random pure states, 
but in general they are not orthogonal. I am going to show that
the detection of entanglement is even more difficult for these
mixed states, as it requires a number of measurements growing
exponentially with $m$.

It is interesting to remark that 
there are other physical contexts in which formally the same 
kind of coarse graining naturally appears:  
\begin{itemize}
\item
(i) {\em Time averaging} - For example, if a state $\ket{\psi}$ undergoes
a time evolution $\ket{\psi(t)} = U(t)\ket{\psi}$ 
given in terms of some unitary dynamics $U(t)$,
then the time average of a physical observable 
$A$ over an interval $T$ is given
by the expectation value ${\rm Tr}(A \rho)$ for the mixed state
\begin{equation}
\rho=\frac{1}{T}\int_0^T {d} t \ket{\psi(t)}\bra{\psi(t)},
\label{eq:rhot}
\end{equation}
which has an effective rank 
$m \approx T/t_{\rm corr}$, where $t_{\rm corr}$ is a dynamical correlation
time of the dynamics $U(t)$. 
For a quantum chaotic evolution $U(t)$, 
the state $\ket{\psi(t)}$ can be, after some time, 
arguably well described by a random state and the correlation 
time $t_{\rm corr}$ is expected to be short,  
so $\rho$ in Eq.~(\ref{eq:rhot}) may be well approximated
by a mixture 
of $m$ uncorrelated random states analogous to Eq.~(\ref{eq:rho}).
\item
(ii) {\em Phase space averaging} - Sometimes it is useful to 
represent quantum states in terms of distribution functions in the 
classical phase space, like the Husimi function
(see, e.g.,~\cite{saraceno}), which can be understood as a 
convolution of the Wigner function or its coarse graining over 
a phase space volume $2\pi\hbar$ (to simplify writing, let us consider
systems with one degree of freedom).
In fact, the Husimi function of a pure state can be understood 
as a Wigner function of the following mixed state:
 \begin{equation}
 \rho = \frac{1}{2\pi\hbar} \int {d}{q}{d}{p}
 \exp\left[-\frac{1}{2\hbar}(\alpha q^2+\alpha^{-1} p^2)\right] 
 T({q},{p})\ket{\psi}\bra{\psi}T^\dagger({q},{p}),
 \label{eq:rhop} \end{equation}
where $T({q},{p})$ are unitary 
phase space translation operators, and $\alpha$ is 
an arbitrary (squeezing) parameter. 
A random  
pure state $\ket{\psi}$ has a Wigner function with random 
sub-Planck structures with phase space correlation length 
$l_{\rm corr} \sim \hbar$ which is semi-classically smaller 
than the coarse-graining width $\sim \hbar^{1/2}$, so 
$\rho$ in Eq.~(\ref{eq:rhop}) can be again considered 
as a mixture  
(Eq.~(\ref{eq:rho})) of $m$ random pure states with $m \sim \hbar^{-1/2}$.
\end{itemize}

\subsection{Unknown random states}

In this section is is assumed that the random state 
$\ket{\psi}$ whose entanglement we would like to detect is unknown so 
that we are not able to use an optimal entanglement witness 
$W$ for a particular $\ket{\psi}$. 
The best one can do is to choose some fixed witness $W$ in advance,
independently of the state. Since I am interested in the average
behavior over unitary invariant ensemble of pure random states, $W$
can be chosen to be random as well. That is, in the present section
I am going to study detection of entanglement with a random
entanglement witness, whose precise definition will be given later.
What I want to calculate is the distribution of the expectation
values $\bracket{\psi}{W}{\psi}$ for a fixed $W$ and an ensemble of
random pure states $\ket{\psi}$. Averaging over random states $\ket{\psi}$ 
we see that the average expectation value
$\overline{\bracket{\psi}{W}{\psi}}$ is
\begin{equation}
\int d{\cal P} 
\bracket{\psi}{W}{\psi}={\rm Tr}(W)/N, 
\end{equation}
where
$\overline{\bullet} = \int d{\cal P}\bullet$ denotes an integration 
over the $U(N)$-invariant (Haar) distribution
of pure states  $\ket{\psi}$, and I used the fact that for
a random state $\ket{\psi}=\sum_i c_i \ket{i}$
we have $\overline{c_i c_j^*}=\delta_{ij}/N$.
Let us fix normalization of the entanglement witness $W$ such that
${\rm Tr}(W)=1$. Therefore, the average expectation value
$\overline{\bracket{\psi}{W}{\psi}}$ scales $\propto 1/N$.
It is therefore convenient to define the rescaled quantity
$w=N \bracket{\psi}{W}{\psi}$ such that $\overline{w}=1$,
independently of the dimension $N$. 

Here and in the following,
the investigation is limited to decomposable entanglement witnesses
(see Appendix~\ref{sec:witness}) of the form $W=Q^{\rm T_B}$,
with $Q$ positive semidefinite operator.

\begin{figure}
\begin{center}
\includegraphics[width=70mm,angle=-90]{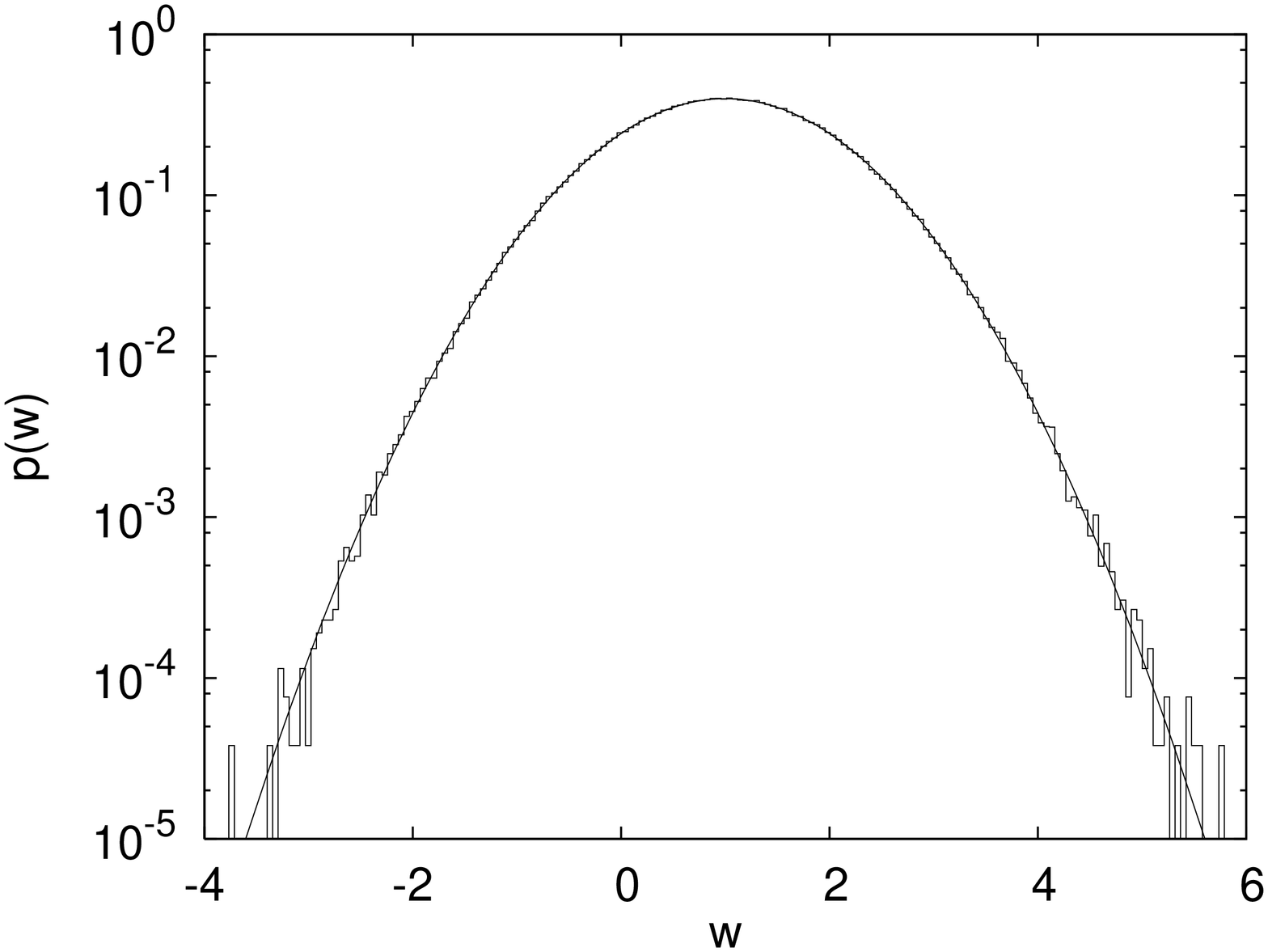}
\\
\includegraphics[width=70mm,angle=-90]{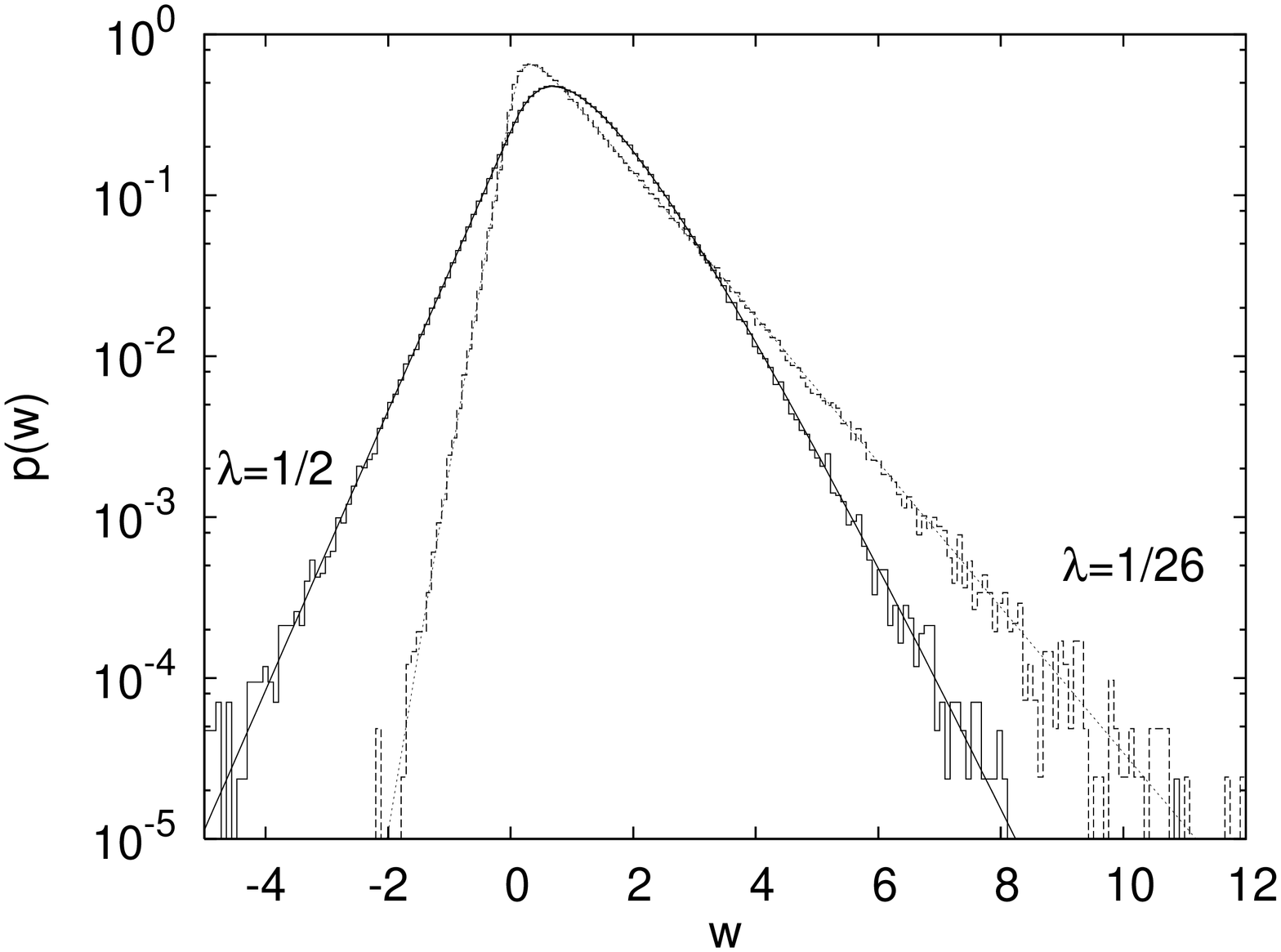}
\end{center}
\caption{Distribution of $w=N\bracket{\psi}{W}{\psi}$ for random
vectors $\ket{\psi}$ and a single $W=(\ket{\phi}\bra{\phi})^{\rm T_B}$
with a random $\ket{\phi}$ (top) and 
with $\ket{\phi}$ having two nonzero Schmidt coefficients
$\sqrt{\lambda}$ and $\sqrt{1-\lambda}$
(two cases are shown, $\lambda=1/2$ and $\lambda=1/26$) (bottom).
Histograms correspond to numerical simulations
for $N=2^{10}$, curves to the theoretical predictions
of Eqs.~(\ref{eq:gauss}) (top) and 
and (\ref{eq:pw_sch}) (bottom).
This figure is taken from Ref.~\cite{slovenia}.}
\label{fig:gauss}
\end{figure}

I first consider the case when $Q$ is a simple 
rank one projector, that is, $W$
is given by $W=(\ket{\phi}\bra{\phi})^{\rm T_B}$. 
If $\ket{\phi}$ is a state with a large Schmidt number
$r\sim \sqrt{N}$, as it is typical for random $\ket{\phi}$
(I consider equal size subsystems, $N_A=N_B=\sqrt{N}$), 
then one can show~\cite{slovenia} that the probability density 
$p(w)=d{\cal P}/dw$ converges to a Gaussian in the limit $N\to \infty$, 
\begin{equation}
p(w)=\frac{1}{\sqrt{2\pi}}\exp{(-(w-1)^2/2)}.
\label{eq:gauss}
\end{equation}
Numerical results for finite $N=2^{10}$ are shown in 
Fig.~\ref{fig:gauss} (top).
The probability of measuring negative $w$, i.e., 
of detecting
entanglement, is $\int_{-\infty}^0{p(w)dw}$ and therefore
\begin{equation}
{\cal P}(w<0)=(1-{\rm erf}(1/\sqrt{2}))/2 \approx 0.159.
\end{equation}
Note that this entanglement detection probability is independent
of the details of $\ket{\phi}$, provided that its Schmidt
number $r$ is large, more precisely $r\propto N$.

Since it appears difficult to measures witness operators 
corresponding to states $\ket{\phi}$ with large Schmidt number $r$, 
it is interesting to consider the opposite limit of small $r$.
In particular, let us consider the extreme case of rank $r=2$.
We therefore have only two nonzero terms in 
the Schmidt decomposition (\ref{schmidtdec}),
corresponding to the nonzero eigenvalues,
$p_1=\lambda$ and $p_2=1-\lambda$, of the reduced density matrix 
$\rho_A={\rm Tr}_{B}\big(|\phi\rangle\langle\phi|\big)$.
In this case, we obtain~\cite{slovenia}
\begin{equation}
p(w)=
\cases{
  \frac{1}{(1-2\lambda)^2} \left\{\lambda {\rm e}^{-\frac{w}{\lambda}} 
+ (1-\lambda){\rm e}^{-\frac{w}{1-\lambda}} \right\}
+ \frac{1}{4\sqrt{\lambda(1-\lambda)}-2}{\rm e}^{-\frac{w}{
\sqrt{\lambda(1-\lambda)}}} & : $w>0$, \cr
  \frac{1}{4\sqrt{\lambda(1-\lambda)}+2}{\rm e}^{\frac{w}{
\sqrt{\lambda(1-\lambda)}}}  & : $w<0$.
}
\label{eq:pw_sch}
\end{equation}
Results of numerical simulation for $p(w)$ for two cases,
$\lambda=1/2$ and $\lambda=1/26$, are compared in 
in Fig.~\ref{fig:gauss} (bottom).
The probability of detecting entanglement, i.e., 
of measuring negative values of $w$ is 
\begin{equation}
{\cal P}(w<0)=1/(4+2/\sqrt{\lambda(1-\lambda)}).
\end{equation}
For instance, ${\cal P}(w<0)=1/8=0.125$ when $\lambda=1/2$ and
${\cal P}(w<0)=5/72\approx 0.07$ when $\lambda=1/26$.
Note that in
the limit $\lambda \to 0$, i.e., of a pure separable state for
$\ket{\phi}$, $w$ is always positive with an exponential distribution.

I emphasize that, neglecting problems related to finite measurement 
resolution (an issue that will be discussed in 
Sec.~\ref{sec:knownrandomstates}), the entanglement detection 
probability is, for pure states, independent of the system size $N$.

So far I have discussed only the case when $Q$ is of rank one,
$Q=\ket{\phi}\bra{\phi}$. In general, one can consider $Q$ of 
rank $k$:
$Q=\sum_i^k d_i \ket{\phi_i}\bra{\phi_i}$. Since I have
fixed ${\rm Tr}(W)=1$, then $\sum_i d_i=1$. 
Assuming for simplicity that all
$d_i$ are the same, $d_i=1/k$, we obtain~\cite{slovenia},
in the limit $N\to\infty$,
\begin{equation}
p(w)=\sqrt{\frac{k}{2\pi}} {\rm e}^{-k(w-1)^2/2}.
\label{eq:pwr}
\end{equation}
Therefore, the entanglement detection for pure states is more 
efficient if one consider $W=Q^{T_B}$, with $Q$ 
rank-one projector.

It is also interesting to consider the case in which $Q$ is of rank $1$
but we wish to detect the entanglement of mixed states, for instance
of a mixture of $m$ pure random states, as given in Eq.~(\ref{eq:rho}).
In this case, 
the resulting distribution $p(w)$ is, in the limit $N\to\infty$, 
a Gaussian of variance $1/m$~\cite{slovenia}.
Because the distribution $p(w)$ becomes narrowly peaked about its
mean $\overline{w}=1$ with increasing $m$, the probability of measuring
negative values decreases with $m$, that is, we obtain
\begin{equation}
{\cal P}(w<0)=\frac{1-{\rm erf}(\sqrt{m/2})}{2}\asymp
\frac{1}{\sqrt{2\pi m}} {\rm e}^{-m/2}.
\label{eq:pm}
\end{equation}
This probability decays to zero exponentially with $m$.
Therefore, the detection of entanglement for a mixture of random states
is very hard.
This result outlines the importance of coarse graining to
explain the emergence of classicality. 
For random pure states, a finite success probability
in the detection of entanglement exists also in the limit in which
the Hilbert space dimension $N\to\infty$. This implies that 
chaotic dynamics alone is not sufficient to erase any trace of entanglement
when going to the classical limit, 
provided that ideal measurements are possible.
On the other hand such erasure becomes
very efficient when coarse graining is taken into account, for
instance when mixtures instead of pure states are considered.

\subsection{Known random states}

\label{sec:knownrandomstates}

In this section it is assumed that the random state $\ket{\psi}$ whose
entanglement we want to measure is known in advance and furthermore,
that we are able to prepare an arbitrary decomposable entanglement witness.
In addition, we have to assume that our state $\ket{\psi}$ is neither 
separable, nor
bound entangled (see Appendix~\ref{app:separability}), 
which is true with probability which converges to 
one exponentially in $N$. Therefore, for
each $\ket{\psi}$ we can prepare an optimal entanglement witness,
such that its expectation value will be minimal. As far as 
decomposable entanglement witnesses
are concerned, the optimal choice of $W=W_{\rm opt}$ is to take for $Q$
a projector to the eigenspace corresponding to the minimal
(negative) eigenvalue $\lambda_{\rm min}$ of $\rho^{\rm T_B}$,
$W_{\rm opt}=(\ket{\phi_{\rm min}}\bra{\phi_{\rm min}})^{\rm T_B}$.
The maximal violation of positivity is therefore
\begin{equation}
{\rm Tr}{(W_{\rm opt}\rho)}=-|\lambda_{\rm min}(\rho^{\rm T_B})|.
\label{eq:Wopt}
\end{equation}
If we are able to measure the 
entanglement witness $W_{\rm opt}$ with a given precision
it is the size of $\lambda_{\rm min}$ which determines the difficulty
of detecting entanglement in $\ket{\psi}$.
Note that the optimal
entanglement witness $W_{\rm opt}$ depends on the state $\ket{\psi}$.
For each state $\ket{\psi}$ we have to pick a different $W_{\rm opt}$.

The expectation value of the minimal eigenvalue equals
$\overline{\lambda}_{\rm min}=-4/\sqrt{N}$~\cite{slovenia}. In fact, 
the distribution of $\lambda_{\rm min}$ becomes
strongly peaked around $-4/\sqrt{N}$ with diminishing relative 
fluctuations as $N\to\infty$.
When we mix several independent (in general non-orthogonal) random vectors,
$\rho=\sum_i^m \ket{\psi_i}\bra{\psi_i}/m$, the minimal eigenvalue
$\lambda_{\rm min}$ increases and the distribution becomes increasingly
sharply peaked (for $m \to \infty$ we get $\rho \to \mathbbm{1}/N$
with all eigenvalues being equal to $1/N$). 
Note that the average minimal eigenvalue $\bar{\lambda}_{\rm min}$ 
is positive for $m > m^*$, with $m^* \approx 4 N$.

Although von Neumann entropy of a random state is large all eigenvalues
of $\rho^{\rm T_B}$ are very small and will therefore be hard to detect.
If we assume that we are able to measure values of $|{\rm Tr}{(W\rho)}| 
< \epsilon$
then we can, depending on the scaling of $\epsilon$ with $N$, tell for which
values of $m$ the detection of entanglement is possible. If $\epsilon$ does not
depend on $N$, i.e., precision does not increase with $N$, then for
sufficiently large $N$, such that $4/\sqrt{N} <\epsilon$, 
detection of entanglement
will be impossible. Already a single random state becomes from the viewpoint
of entanglement detection ``classical'', since 
measuring a negative expectation value
of its optimal entanglement witness is below the detection limit. If on
the other hand we are able to measure $\epsilon$ which decreases as
$1/\sqrt{N}$, the critical $m_{\rm crit}$, beyond which the entanglement
detection is impossible, will be independent of $N$, i.e., in the
limit $N\to\infty$ the ratio $m_{\rm crit}/\sqrt{N} \to 0$. 
If however we are able
to detect very small expectation values of order $1/N$, then
$m_{\rm crit}$ will be proportional to $N$.
Furthermore, even with arbitrary accuracy, detection of entanglement 
with decomposable entanglement witnesses is in practice impossible
beyond $m=m^* \propto N$ (that is, when $\bar{\lambda}_{\rm min}$
becomes positive).

I would like to stress once more than the detection
difficulties are a consequence of the complexity of random states.
If instead one considers ``regular states'' such as the 
GHZ state~\cite{GHZstate}
$|{\rm GHZ}\rangle=\frac{1}{\sqrt{2}}
(|0...0\rangle+|1...1\rangle)$, then the optimal
witness is 
$W_{\rm opt}=(|\phi_{\rm min}\rangle\langle\phi_{\rm min}|)^{\rm T_B}$,
with $|\phi_{\rm min}\rangle=
\frac{1}{\sqrt{2}}(|0...0\rangle|1...1\rangle
-|1...1\rangle|0...0\rangle)$ which corresponds to the minimal eigenvalue
$\lambda_{\rm min}=-1/2$ of $(|{\rm GHZ}\rangle
\langle{\rm GHZ}|)^{\rm T_B}$.
Since the value of $\lambda_{\rm min}$ is $-1/2$ instead of $-4/\sqrt{N}$
as typical for a random state, it turns out that it will be much easier
to detect entanglement in a ``regular'' rather than in a random
state. This happens in spite of the fact that the entanglement
content is much larger in a random than in such a regular state.

\section{Chaotic environments}

\label{sec:chaoticenvironments}

Real physical systems are never isolated and the coupling of the
system to the environment leads to decoherence. This process can
be understood as the loss of quantum information, initially
present in the state of the system, when non-classical
correlations (entanglement) establish between the system and
the environment. On the other
hand, when tracing over the environmental degrees of freedom,
we expect that the entanglement between internal degrees
of freedom of the system is reduced or even destroyed.
Decoherence theory has a fundamental interest,
since it provides explanations of the emergence of classicality
in a world governed by the laws of quantum mechanics~\cite{zurek}.
Moreover, it is a threat to the
actual implementation of any quantum computation and
communication protocol~\cite{qcbook,nielsen}.
Indeed, decoherence invalidates the
quantum superposition principle, which is at the heart of the
power of quantum algorithms. 
A deeper understanding
of the decoherence phenomenon is essential to develop
quantum technologies.

The environment is usually described as a many-body quantum system.
The best-known model is the Caldeira-Leggett 
model~\cite{caldeiraleggett,ingold,weiss},
in which the environment is a bosonic bath consisting of infinitely
many harmonic oscillators at thermal equilibrium.
More recently, first studies of the role played by chaotic
dynamics~\cite{park,blume03,lee,saraceno06,rossini06,seligman}
or random environments~\cite{Pineda2007,Petruccione2007}
in the decoherence process have been carried out. 

In the following, it is shown that 
the many-body environment may be substituted with a closed deterministic
system with a small number of degrees of freedom, 
but chaotic~\cite{rossini06}. In
other words, the complexity of the environment arise not from being
many-body but from having chaotic dynamics.
I consider two qubits coupled to a 
\emph{single particle, fully deterministic, conservative
chaotic ``environment''}, described by the kicked rotator model.
It is shown that, due to the system-environment interaction,
the entropy of the system increases. At the same time, the
entanglement between the two qubits decays, thus illustrating
the loss of quantum coherence. The evolution in
time of the two-qubit entanglement is in good agreement with the
evolution obtained in a pure dephasing stochastic model.
Since this pure dephasing decoherence mechanism can be derived
in the framework of the Caldeira-Leggett model~\cite{palma},
a direct link between the effects
of a many-body environment and of a chaotic single-particle
environment is established.

Let us consider two qubits coupled to a
quantum kicked rotator. The overall system is
governed by the Hamiltonian
\begin{equation}
\hat{H} = \hat{H}^{(1)} + \hat{H}^{(2)} +
\hat{H}^{(\mathrm{kr})} + \hat{H}^{(\mathrm{int})} \, ,
\label{eq:hammodel} 
\end{equation}
where $\hat{H}^{(i)} = \omega_i \, \hat{\sigma}_x^{(i)}$ ($i=1,2$)
describes the free evolution of the two qubits,
\begin{equation}
\hat{H}^{(\mathrm{kr})} = \frac{\hat{n}^2}{2} + k \cos(\hat{\theta})
\sum_j \delta(\tau-jT)
\label{eq:krotator}
\end{equation}
the quantum kicked rotator, and
\begin{equation}
\hat{H}^{(\mathrm{int})} = \epsilon \: ( \hat{\sigma}_z^{(1)} +
\hat{\sigma}_z^{(2)}) \cos(\hat{\theta}) \sum_j \delta(\tau-jT)
\end{equation}
the interaction between the qubits and the kicked rotator;
as usual, 
$\hat{\sigma}_\alpha^{(i)}$ ($\alpha = x,y,z$) denote the Pauli 
operators for the $i$-th qubit.
Both the cosine potential in $\hat{H}^{(\mathrm{kr})}$ and
the interaction $\hat{H}^{(\mathrm{int})}$ are switched on and off
instantaneously (kicks) at regular time intervals $T$.
Let us consider the two qubits as an open quantum system and the kicked
rotator as their \emph{common} environment.
Note that I chose non-interacting qubits as I want their entanglement
to be affected exclusively by the coupling to the 
environment~\footnote{As discussed in~\cite{lee}, 
the chaotic environment model discussed in this 
section could be implemented, at least in principle, using cold atoms
in a pulsed optical lattice created by laser fields~\cite{darcy} or
superconducting nanocircuits~\cite{romito}.}

The kicked rotator, as the sawtooth map described in detail in 
Appendix~\ref{sec:sawmap},
belongs to the class of periodically driven systems
of Eq.~(\ref{sawham}), with the external driving 
described by the potential $V(\theta)=k \cos\theta$,
switched on and off instantaneously at time intervals $T$.
The evolution from time $tT^-$ (prior to the $t$-th kick) to time $(t+1)T^-$
(prior to the $(t+1)$-th kick) of the kicked rotator in the classical limit
is described by the Chirikov standard map:
\begin{equation}
\left\{ 
\begin{array}{l}
n_{t+1} = n_t + k \sin \theta_t, \\
\theta_{t+1} = \theta_t + T n_{t+1},
\end{array} 
\right. 
\label{eq:chirikov}
\end{equation}
where $(n,\theta)$ are conjugated momentum-angle variables and
$t=\tau/T$ denotes the discrete time, measured in number of kicks.
As for the sawtooth map, 
by rescaling $n \to p = T n$, the dynamics of Eq.~(\ref{eq:chirikov})
is seen to depend only on the parameter $K = kT$.
For $K=0$ the motion is integrable; when $K$ increases,
a transition to chaos of the Kolmogorov-Arnold-Moser (KAM) 
type is observed~\cite{lichtenberg,arnold}.~\footnote{The property of 
complete integrability is very delicate and atypical
as it is, in general, destroyed by an arbitrarily weak perturbation
that converts a completely integrable system into a KAM-integrable
system. The structure of KAM 
motion is very intricate: the
motion is confined to invariant tori for most initial conditions yet a single,
connected, chaotic motion component (for more than two degrees of freedom) 
of exponentially
small measure (with respect to the perturbation) arises, which is nevertheless
everywhere dense.} 
The last invariant KAM torus is broken for $K \approx  0.97$.
If $K \sim 1$ the phase space is mixed (simultaneous presence of
integrable and chaotic components).
If $K$ increases further, the stability islands progressively reduce
their size; for $K \gg 1$ they are not visible any more.
In what follows, I always consider map (\ref{eq:chirikov}) on the torus
$0 \leq \theta < 2 \pi$, $- \pi \leq p < \pi$.
In this case, the Chirikov standard map describes the stroboscopic dynamics of
a \emph{conservative} dynamical system with two degrees of freedom which,
in the fully chaotic regime $K\gg 1$, relaxes, apart from quantum fluctuations,
to the uniform distribution on the torus.

The Hilbert space of the global system is given by
\begin{equation}
\mathcal{H} = \mathcal{H}^{(1)} \otimes \mathcal{H}^{(2)} \otimes
\mathcal{H}^{(kr)} \, ,
\end{equation}
where $\mathcal{H}^{(1)}$ and $\mathcal{H}^{(2)}$ are the two-dimensional
Hilbert spaces associated to the two qubits, and
$\mathcal{H}^{(kr)}$ is the Hilbert space for the kicked rotator
with $N$ quantum levels.

The time evolution generated by Hamiltonian
(\ref{eq:hammodel}) in one kick is described by the operator
\begin{equation}
\begin{array}{rl}
\hat{U} = & \exp \big[  -i \big( k + \epsilon (\hat{\sigma}_z^{(1)} +
\hat{\sigma}_z^{(2)}) \big) \cos(\hat{\theta}) \big]  \\
& \times\, \exp \big[ -i T \frac{\hat{n}^2}{2} \big]
\exp \big( -i \, \delta_1 \, \hat{\sigma}_x^{(1)} \big)
\exp \big( -i \, \delta_2 \, \hat{\sigma}_x^{(2)} \big).
\end{array} \label{eq:kickedevol}
\end{equation}
The effective Planck constant is $\hbar_{\rm eff}=T = 2 \pi /N$;
$\delta_1 = \omega_1 T , \: \delta_2 = \omega_2 T$;
$\epsilon$ is the coupling strength between the qubits and the environment.
The classical limit $\hbar_{\rm eff} \to 0$ is obtained by taking
$T \to 0$ and $k \to \infty$, in such a way that $K = kT$ is kept fixed.

I am interested in the case 
in which the environment (the kicked rotator) is chaotic
(that is, with $K \gg 1$).
The two qubits are initially prepared in a maximally entangled state,
so that they are disentangled from the environment.
Namely, I suppose that at $t=0$ the system is in the state
\begin{equation}
\ket{\Psi_0} = \ket{\phi^+} \otimes \ket{\psi_0} \, ,
\label{eq:initial}
\end{equation}
where $\ket{\phi^+} = \frac{1}{\sqrt{2}} \left( \ket{00} + \ket{11} \right)$
is a Bell state (the particular choice of the initial maximally entangled
state is not crucial for what follows), and
$\ket{\psi_0} = \sum_n c_n \ket{n}$ is a generic state of
the kicked rotator, with $c_n$ random coefficients such that
$\sum_n |c_n|^2 =1$, and $\ket{n}$ eigenstates of the momentum
operator.
The evolution in time of the global system (kicked rotator plus qubits)
is described by the unitary operator 
$\hat{U}$ defined in Eq.~(\ref{eq:kickedevol}).
Therefore, any initial pure
state $\ket{\Psi_0}$ evolves into another pure state
$\ket{\Psi (t)} = \hat{U}^t \ket{\Psi_0}$.
The reduced density matrix $\rho_{12} (t)$ describing the two qubits
at time $t$ is then obtained after tracing $\ket{\Psi (t)} \bra{\Psi (t)}$
over the kicked rotator's degrees of freedom.

In the following I will focus my attention on the time evolution
of the \emph{entanglement of formation} (see Appendix~\ref{sec:concurrence})
$E_{12}$ between the two qubits
and that between them and the kicked rotator, measured by 
the reduced von Neumann entropy 
$S_{12}=-\mathrm{Tr} \, [ \rho_{12}\log_{2} \rho_{12}]$
of the reduced density matrix $\rho_{12}$.
Clearly, for states like the one in Eq.~(\ref{eq:initial}),
we have $E_{12} (0) = 1$, $S_{12} (0) = 0$.
As the total system evolves, we expect that $E_{12}$ decreases, while
$S_{12}$ grows up, thus meaning that the two-qubit system is
progressively losing coherence.

If the kicked rotator is in the chaotic regime and in the semiclassical
region $\hbar_{\rm eff}\ll 1$, it is possible to drastically
simplify the description of the system in Eq.~(\ref{eq:hammodel})
by using the \emph{random phase-kick} approximation, in the framework
of the Kraus representation formalism.
Since, to a first approximation, the phases between two consecutive kicks
in the chaotic regime can be considered as uncorrelated, the interaction
with the environment can be simply modeled as a phase-kick rotating
both qubits through the same random angle about the $z$-axis
of the Bloch sphere.
This rotation is described in the 
$\{|00\rangle,|01\rangle,|10\rangle,|11\rangle\}$ basis by 
the unitary matrix
\begin{equation} \label{eq:rotation}
{R} (\theta) = \left[
\begin{array}{cc} e^{- i \epsilon \cos \theta} & 0 \\
0 & e^{i \epsilon \cos \theta} \end{array} \right] \otimes
\left[ \begin{array}{cc} e^{- i \epsilon \cos \theta} & 0 \\
0 & e^{i \epsilon \cos \theta} \end{array} \right],
\end{equation}
where the angle $\theta$ is drawn from a uniform random
distribution in $[ 0, 2\pi )$.
The one-kick evolution of the reduced density matrix $\rho_{12}$ is then
obtained after averaging over $\theta$:
\begin{equation}
\bar{\rho}_{12} = \frac{1}{2\pi} \int_{0}^{2 \pi} \hspace{-2mm} d \theta \,
{R} (\theta) \, e^{-i \delta_2 {\sigma}_x^{(2)}}
e^{-i \delta_1 {\sigma}_x^{(1)}}
\,\rho_{12} \,
e^{i \delta_1 {\sigma}_x^{(1)}} e^{i \delta_2 {\sigma}_x^{(2)}}
{R}^{\dagger} (\theta). \label{eq:randomphase}
\end{equation}

\begin{figure}
\begin{center}
\includegraphics[width=10.cm]{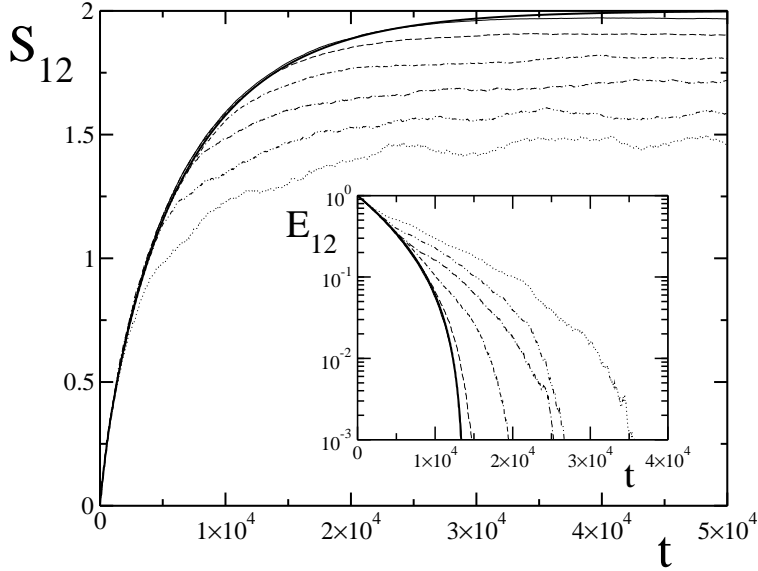}
\caption{Reduced von Neumann entropy $S_{12}$ (main figure) and
entanglement $E_{12}$ (inset) as a function of time at $K \approx 99.73$,
$\delta_1=10^{-2}$, $\delta_2=\sqrt{2}\delta_1$,
$\epsilon=8\times 10^{-3}$. The thin curves correspond
to different number of levels for the environment (the kicked rotator)
($N=2^9,2^{10},2^{11},2^{12},2^{13},2^{14}$ from bottom to
top in the main figure and vice versa in the inset).
The thick curves give the numerical results from the random phase model
(\ref{eq:randomphase}).}
\label{fig:confr_ES}
\end{center}
\end{figure}

In order to assess the validity of the random phase-kick approximation,
model (\ref{eq:hammodel}) is numerically investigated in the
classically chaotic regime $K\gg 1$ and in the
region $\hbar_{\rm eff}\ll 1$ in which the environment is a semiclassical
object. Under these conditions, we expect that the
time evolution of the entanglement can be accurately predicted
by the random phase model. Such expectation is confirmed by the numerical
data shown in  Fig.~\ref{fig:confr_ES}.
Even though differences between the
two models remain at long times due to the finite number $N$ of
levels in the kicked rotator, such differences appear at later
and later times when $N\to \infty$ ($\hbar_{\rm eff}\to 0$).
The parameter $K$ has been chosen much greater than one,
so that the classical phase space of the
kicked rotator can be considered as completely chaotic.
Note that the value $K \approx 99.72676$ is chosen 
to completely wipe off
memory effects between consecutive and next-consecutive kicks
(see Ref.~\cite{rossini06} for details).

I point out that the random phase model can be derived from 
the Caldeira-Leggett
model with a pure dephasing coupling
$\propto ( \hat{\sigma}_z^{(1)} + \hat{\sigma}_z^{(2)}) \sum_k g_k \hat{q}_k$,
with $g_k$ coupling constant to the 
$k$-th oscillator of the environment, whose coordinate
operator is $\hat{q}_k$~\cite{palma,braun}. This
establishes a direct link between the chaotic single-particle environment
considered in this paper and a standard many-body environment.

\section{Final remarks}

The role of entanglement as a resource in quantum information has 
stimulated intensive research aimed at unveiling both its qualitative 
and quantitative aspects. The interest is first of all motivated 
by experimental implementations of quantum information protocols.
Decoherence, which can be considered as the ultimate obstacle in the way 
of actual implementation of any quantum computation or communication 
protocol, is due to the entanglement between the quantum hardware 
and the environment. 
The decoherence-control issue is expected to be particularly relevant 
when the state of the quantum system is \emph{complex}, namely when 
it is characterized by a large amount of multipartite entanglement. 
It is therefore important, for applications but also in its own right, 
to scrutinize the robustness and the multipartite features of  
relevant classes of entangled states. 
In this context, random states play an important role, both for 
applications in quantum protocols and in view of a, highly desirable,
statistical theory of entanglement.

Such studies have deep links with the physics of complex systems.
In classical physics, a well defined notion of complexity, based 
on the exponential instability of chaos, exists, and has profound 
links with the notion of algorithmic complexity~\cite{ford,alekseev}:
in terms of the symbolic dynamical description, almost all orbits are 
random and unpredictable. 
On the other hand, in spite of many efforts (see~\cite{prosen2007}
and references therein)
the transfer of these concepts to quantum mechanics still remains 
elusive. However, there is strong numerical evidence that 
quantum motion is characterized by a greater degree of 
stability than classical motion (see~\cite{arrow,varenna05,qcbook,sokolov}). 
This has important consequences on the stability of quantum algorithms;
for instance, the robustness of the multipartite entanglement 
generated by chaotic maps and discussed in 
Sec.~\ref{sec:stabilitymultipartite} is related to the 
power-law decay of the fidelity time scales 
for quantum algorithms~\cite{georgeot2000,thermalization,benenti01,frahm}
which, in turn, is a consequence of the discreteness of 
the phase space in quantum mechanics~\cite{arrow,varenna05,qcbook,sokolov}.
If we consider the chaotic
classical motion (governed by the Liouville equation) of some phase-space 
density, smaller and smaller scales are explored exponentially fast.
These fine details of the density distribution are rapidly lost under small
perturbations. In quantum mechanics, there is a lower limit to this process, set
by the size of the Planck cell, and this reduces the complexity of 
quantum motion as compared to classical motion.

Finally, the fundamental, purely quantum notion of entanglement is expected 
to play a crucial role in characterizing 
the complexity of a quantum system~\cite{briegel2005}. 
I believe that studies of complexity and multipartite entanglement
will shed some light on series of 
very important issues in quantum computation and in critical 
phenomena of quantum many-body condensed matter physics. 

\appendix

\section{Separability criteria}

\label{app:separability}

\subsection{The Peres criterion}

\label{app:peres}

The Peres criterion~\cite{peres96} provides a necessary condition for the
existence of decomposition (\ref{sepdecomposition}), in
other words, a violation of this criterion is a sufficient condition for
entanglement. This criterion is based on the
\emph{partial transpose} operation.
Introducing an orthonormal basis 
$\{|i\rangle_{A}|\alpha\rangle_{B}\}$ in
the Hilbert space $\mathcal{H}_{AB}$ associated with the bipartite system $A+B$,
the density matrix $\rho_{AB}$ has matrix elements $(\rho_{AB})_{i\alpha;j\beta}
= {}_{A}\langle i|{}_B\langle\alpha|
\rho_{AB}|j\rangle_{A}|\beta\rangle_{B}$. The
partial transpose density matrix is constructed by only 
taking the transpose in either the Latin or Greek
indices (here Latin indices
refer to Alice's subsystem and Greek indices to Bob's). 
For instance, the partial transpose with respect to Alice is given
by
\begin{equation}
  \big(\rho^{T_A}_{AB}\big)_{i\alpha;j\beta} =
  \big(\rho_{AB}\big)_{j\alpha;i\beta}.
\end{equation}
Since a separable state $\rho_{AB}$ can always be written in the form
(\ref{sepdecomposition}) and the density matrices $\rho_{Ak}$ and $\rho_{Bk}$
have non-negative eigenvalues, then the overall density matrix
$\rho_{AB}$ also has non-negative eigenvalues. The partial transpose of a
separable state reads
\begin{equation}
  \rho^{T_A}_{AB} = \sum_k p_k \, \rho^T_{Ak} \otimes \rho_{Bk}.
\end{equation}
Since the transpose matrices $\rho^T_{Ak}=\rho^\star_{Ak}$ are Hermitian
non-negative matrices with unit trace, they are also legitimate density
matrices for Alice. It follows that none of the eigenvalues of $\rho^{T_A}_{AB}$
is non-negative. This is a necessary condition for decomposition
(\ref{sepdecomposition}) to hold. It is then sufficient to have at least one
negative eigenvalue of $\rho^{T_A}_{AB}$ to conclude that the state $\rho_{AB}$
is entangled.

It can be shown~\cite{horodecki96} that for composite states of dimension
$2\times2$ and $2\times3$, the Peres criterion provides a
necessary and sufficient condition for separability; that is, the state
$\rho_{AB}$ is separable if and only if $\rho^{T_A}_{AB}$ is non-negative.
However, for higher dimensional systems, states exist
for which all eigenvalues of the partial transpose are
non-negative, but that are non-separable~\cite{horodecki97}.
These states are known as
\emph{bound entangled states}
since they cannot be distilled by means of local operations and classical
communication to form a maximally entangled state~\cite{horodecki98}.

I stress that the Peres criterion is more sensitive than Bell's inequality for
detecting quantum entanglement; that is, there are states detected as
entangled by the Peres criterion that do not violate Bell's
inequalities (\cite{peres96}).

\subsection{Entanglement witnesses}

\label{sec:witness}

A convenient way to detect entanglement is to use
the so-called entanglement witnesses~\cite{horodecki96,terhal00}.
By definition, an entanglement witness is a Hermitian operator $W$
such that ${\rm Tr}\big(W\rho^{(\rm sep)}_{AB}\big)\geq 0$ 
for all separable states $\rho^{(\rm sep)}_{AB}$
while there exists at least one state $\rho^{(\rm ent)}_{AB}$ such that
${\rm Tr}\big(W\rho^{(\rm ent)}_{AB})<0$. Therefore, 
the negative expectation value
of $W$ is a signature of entanglement and the state 
$\rho^{(\rm ent)}_{AB}$
is said to be detected as entangled by the witness $W$.

The existence of entanglement witnesses is a consequence of the 
\emph{Hahn-Banach theorem:} Given a convex and compact set $S$ and
$\rho_{AB}\notin S$, there exists a hyperplane that separates 
$\rho_{AB}$ from $S$. 

\begin{figure}
  \begin{center}
    \includegraphics[width=10.0cm]{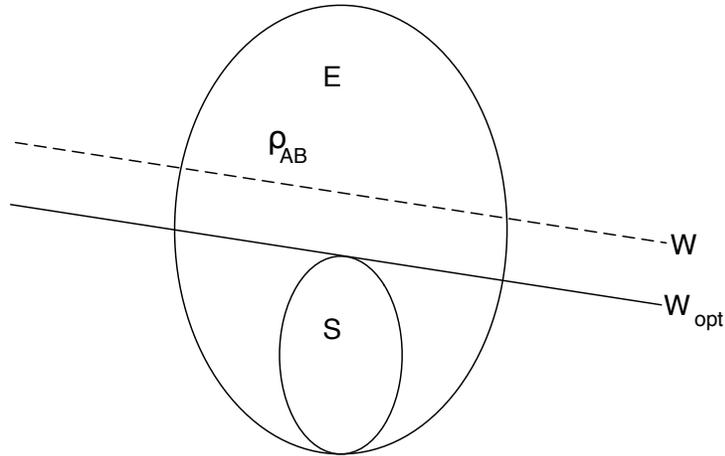}
    \caption{Schematic drawing of entanglement witnesses.}
    \label{fig:witness}
  \end{center}
\end{figure}

This fact is illustrated in Fig.~\ref{fig:witness}.
The set $S$ of separable states is a subset of the set
of all possible density matrices for a given system. The dashed line
represents a hyperplane separating an entangle state $\rho_{AB}$ from
$S$. The optimized witness $W_{\rm opt}$ (represented by a full 
line) is obtained after performing a parallel transport of the 
above hyperplane, so that it becomes tangent to the set of separable 
states. Therefore, the optimized witness $W_{\rm opt}$ detects more 
entangled states than before parallel transport.  
Note that, in order to fully characterize the set $S$ of separable states 
one should find all the witnesses tangent to $S$.

The concept of entanglement witness is close to experimental
implementations and detection of entanglement by means of
entanglement witnesses has been realized in several
experiments~\cite{Bour:04,wineland,Haffner:05}.
The more negative expectation value of entanglement witness we find,
the easier it is to detect entanglement of such a state.
The expectation value of $W$ also provides lower bounds to various entanglement
measures~\cite{eisert:07,Guhne:07}.
Finally, it is interesting to note that violation of Bell's inequalities
can be rewritten in terms of non-optimal entanglement
witnesses~\cite{terhal00,Hyllus:05}.

In general, classification of entanglement witnesses is a hard problem. 
However, much simpler is the issue with the so-called decomposable 
entanglement witnesses~\cite{Lewenstein:00}. 
By definition, a witness is called decomposable if  
\begin{equation}
W=P+Q^{\rm T_B}, \qquad P,Q\ge 0,
\label{eq:DEW}
\end{equation}
that is, with positive semidefinite operators $P,Q$. 
Decomposable entanglement witnesses can only detect 
entangled states with at least one negative eigenvalues
of $\rho^{\rm T_B}$. Therefore, similarly to the Peres criterion,
decomposable witnesses do not detect bound entangled states.
Note, however, that entanglement witnesses are 
closer to experimental implementations than the Peres criterion, 
as full tomographic knowledge about the state is not needed. 

\section{Purity of random states}

\label{app:lubkin}

Let us write a $N$-level random state in the form
\begin{equation}
|\psi\rangle = 
\sum_{k=0}^{N-1} r_k e^{i\phi_k} |k\rangle,
\end{equation}
where $\phi_k$ are independent random variables 
uniformly distributed in $[0,2\pi)$ and 
${\bf r}=(r_0,...,r_{N-1})$ is a random point uniformly 
distributed on the unit hypersphere $\mathbb{S}^{N-1}=
\{{\bf r}\in \mathbb{R}^N \vert {\bf r}^2=1\}$,
with distribution function
\begin{equation}
p({\bf r})=C_N \prod_{k=0}^{N-1} r_k \delta({\bf r}^2-1),
\end{equation}
with $C_N$ normalization constant to be determined later.

Given a bipartition of the Hilbert space of the system into two parts, 
$A$ and $B$, with dimensions $N_A$ and $N_B$, the purity reads
\begin{equation}
P=\sum_{j,j'=0}^{N_A-1} \sum_{l,l'=0}^{N_B-1} 
r_{jl}r_{j'l}r_{j'l'}r_{jl'}
\exp[i(\phi_{jl}-\phi_{j'l}+\phi_{j'l'}-\phi_{jl'})],
\end{equation}
where $|k\rangle=|jl\rangle=|j\rangle_A\otimes |l\rangle_B$.
Following~\cite{facchi}, I split $P$ in two parts:
\begin{equation}
P=X+M,
\end{equation}
where
\begin{equation}
X=\sum_{j,j'}{}^\prime \sum_{l,l'}{}^\prime
r_{jl}r_{j'l}r_{j'l'}r_{jl'}
\exp[i(\phi_{jl}-\phi_{j'l}+\phi_{j'l'}-\phi_{jl'})],
\label{eq:X}
\end{equation}
\begin{equation}
M=\sum_{j,j'}{}^\prime\sum_l r_{jl}^2r_{j'l}^2
+\sum_j \sum_{l,l'}{}^\prime r_{jl}^2r_{jl'}^2
+\sum_{j,l} r_{jl}^4,
\label{eq:M}
\end{equation}
where $\sum'$ means that equal indexes are banned in the sum.
Since $\langle e^{i\phi_k} \rangle =0$ we obtain $\langle X \rangle =0$.
Therefore, 
\begin{equation}
P=\langle M \rangle =
N(N_A+N_B-2) \langle r_0^2 r_1^2 \rangle +
N \langle r_0^4 \rangle,
\label{eq:PM}
\end{equation}
where I have used $\langle r_k^4 \rangle =\langle r_0^4 \rangle$
for all $k$ and $\langle r_k^2 r_{k'}^2 \rangle =
\langle r_0^2 r_1^2 \rangle$ for all $k, k'$ with $k\ne k'$.

I now evaluate the marginal distribution
$$
p(r_0,...,r_{m-1})=C_N
r_0\cdots r_{m-1}
\int_0^1 dr_m r_m 
\int_0^1 dr_{m+1} r_{m+1} 
\cdots 
\int_0^1 dr_{N-1} r_{N-1} \delta (r^2-1)
$$
$$
=\frac{C_N}{2}\,
r_0\cdots r_{m-1}
\int_0^{\sqrt{1-r_0^2-...-r_{m-1}^2}} dr_{m}r_{m}
\cdots
\int_0^{\sqrt{1-r_0^2-...-r_{N-3}^2}} dr_{N-2}r_{N-2}
$$
\begin{equation}
=\frac{C_N}{2^{N-m}}\frac{1}{(N-m-1)!}\,
r_0\cdots r_{m-1}
\left(1-\sum_{j=0}^{m-1} r_j^2\right)^{N-m-1}.
\end{equation}
In particular, 
\begin{equation}
p(r_0)=\frac{C_N}{2^{N-1}}\,\frac{1}{(N-2)!}\,
r_0(1-r_0^2)^{N-2}.
\end{equation}
The normalization condition $\int_0^1 dr_0 p(r_0)=1$ allows us
to determine $C_N=2^N(N-1)!$.
Thus, we obtain
\begin{equation}
p(r_0)=2(N-1)r_0(1-r_0^2)^{N-2},
\end{equation}
\begin{equation}
\langle r_0^4 \rangle 
=\int_0^1 dr_0 r_0^4 p(r_0) =\frac{2}{N(N+1)},
\label{eq:pr0}
\end{equation}
\begin{equation}
p(r_0,r_1)=4(N-1)(N-2)r_0r_1(1-r_0^2-r_1^2)^{N-3},
\end{equation}
\begin{equation}
\langle r_0^2 r_1^2 \rangle 
=\int_0^1 dr_0 r_0^2 \int_0^1 dr_1 r_1^2
p(r_0,r_1) =\frac{1}{N(N+1)}.
\label{eq:pr0pr1}
\end{equation}
After substitution of Eqs.~(\ref{eq:pr0}) and 
(\ref{eq:pr0pr1}) into (\ref{eq:PM}) we readily obtain Lubkin's
formula (\ref{eq:lubkin}).  

The variance $\sigma_P^2$ can be computed with the same 
technique as above~\cite{facchi,caves}. However, to obtain 
the variance (\ref{eq:variancelubkin}) for large $N$
it is suffcient to replace in Eqs.~(\ref{eq:X}) and (\ref{eq:M})
$r_k$ with its mean value $1/\sqrt{N}$:
\begin{equation}
\sigma_P^2=\langle P^2 \rangle -P_L^2=
\langle X^2 \rangle +\langle M^2 \rangle -P_L^2
\approx \langle X^2 \rangle \approx \frac{2}{N^2}.
\end{equation}
We can see from Eqs.~(\ref{eq:X}) and (\ref{eq:M}) that 
$X$ and $M$ are sums of $O(N^2)$ terms of order $1/N^2$.
Therefore, the central limit theorem implies that, for large $N$,
the purity tends to a Gaussian distribution with mean $P_L$
and variance $\sigma_P$.
Finally, I note that all moments of the purity have been recently 
computed~\cite{giraud}.

\section{The sawtooth map}
\label{sec:sawmap}

The sawtooth map is a prototype model in the studies of classical
and quantum dynamical systems and exhibits a rich variety of
interesting physical phenomena, from complete chaos to complete integrability,
normal and anomalous diffusion, dynamical localization, and cantori
localization. Furthermore, the sawtooth map gives a good approximation
to the motion of a particle bouncing inside a stadium billiard (which
is a well-known model of classical and quantum chaos).

\subsection{Classical dynamics}

The sawtooth map belongs to the class of
periodically driven dynamical
systems, governed by the Hamiltonian
\begin{equation}
  H(\theta,n;\tau) =
  \frac{n^2}{2} + V(\theta)
  \sum_{j=-\infty}^{+\infty} \delta(\tau-jT) \,,
  \label{sawham}
\end{equation}
where $(n,\theta)$ are conjugate action-angle variables
($0\leq\theta<2\pi$). This Hamiltonian is the sum of two terms,
$H(\theta,n;\tau)=H_0(n)+U(\theta;\tau)$, where $H_0(n)=n^2\!/2$
is just the kinetic energy of a free rotator (a particle moving on
a circle parametrized by the coordinate $\theta$), while
\begin{equation}
  U(\theta;\tau) =
  V(\theta) \sum_j \delta(\tau-jT)
\end{equation}
represents a force acting on the particle
that is switched on and off instantaneously at time intervals
$T$. Therefore, we say that the dynamics described by Hamiltonian
(\ref{sawham}) is \emph{kicked}.
The corresponding Hamiltonian
equations of motion are
\begin{equation}
  \left\{
    \begin{array}{l}
      \displaystyle
      \dot{n} =
      -\frac{\partial{H}}{\partial\theta} =
    -\frac{d V(\theta)}{d \theta}
    \sum_{j=-\infty}^{+\infty} \delta(\tau-jT) \,,
    \\[2ex]
      \displaystyle
      \dot\theta =
      \frac{\partial{H}}{\partial{n}} = n \,.
    \end{array}
  \right.
\end{equation}
These equations can be easily integrated and one finds that the evolution from
time $tT^-$ (prior to the $t$-th kick) to time $(t+1)T^-$
(prior to the $(t+1)$-th kick) is described by the map
\begin{equation}
  \left\{
    \begin{array}{l}
      \displaystyle
       {n}_{t+1} = n_t + F (\theta) \,,
    \\[2ex]
      \displaystyle
      \theta_{t+1}= \theta_t + T{n}_{t+1} \,,
    \end{array}
  \right.
  \label{sawmap}
\end{equation}
where the discrete time $t=\tau/T$ measures the number of map
iterations and
$F(\theta)=-dV(\theta)/d\theta$ is the force acting on
the particle.

In the following, we focus on the special case
$V(\theta)=-k(\theta-\pi)^2/2$.
This map is called the \emph{sawtooth map}, since the force
$F(\theta)=-dV(\theta)/d\theta=k(\theta-\pi)$ has a sawtooth shape,
with a discontinuity at $\theta=0$. 
For such a discontinuous map the conditions of the
Kolmogorov-Arnold-Moser (KAM) theorem
are not satisfied
and, for any $k\ne 0$, the motion is not bounded
by KAM tori.
By rescaling $n\to{I=Tn}$,
the classical dynamics is seen to depend only on the
parameter $K=kT$. Indeed, in terms of the variables $(I,\theta)$ map
(\ref{sawmap}) becomes
\begin{equation}
  \left\{
    \begin{array}{l}
      \displaystyle
      {I}_{t+1} = I_t + K (\theta - \pi) \,,
    \\[2ex]
      \displaystyle
      {\theta}_{t+1} = \theta_t + {I}_{t+1} \,.
    \end{array}
  \right.
  \label{sawmap2}
\end{equation}
The sawtooth map exhibits sensitive dependence
on initial conditions, which is the
distinctive feature of classical chaos: any small error
is amplified exponentially in time. In other
words, two nearby trajectories separate exponentially, with a rate given
by the maximum Lyapunov exponent $\lambda$, defined as
\begin{equation}
  \lambda =
  \lim_{|t|\to\infty} \frac1{t}
  \ln\!\left( \frac{\delta_t}{\delta_0} \right) ,
\end{equation}
where $\delta_t=\sqrt{[\delta I_t]^2+[\delta \theta_t]^2}$. To
compute $\delta I_t$ and $\delta \theta_t$, we differentiate map
(\ref{sawmap2}), obtaining
\begin{equation}
  \left[
    \begin{array}{c}
      \delta {I}_{t+1} \\
      \delta \theta_{t+1}
    \end{array}
  \right]
  = M
  \left[
    \begin{array}{c}
      \delta I_t \\
      \delta\theta_t
    \end{array}
  \right]
  =
  \left[
    \begin{array}{c@{\quad}c}
      1 &  K  \\
      1 & 1+K
    \end{array}
  \right]
  \left[
    \begin{array}{c}
      \delta I_t \\
      \delta\theta_t
    \end{array}
  \right] .
\label{tangmap}
\end{equation}
The iteration of map (\ref{tangmap}) gives $\delta I_t$ and
$\delta \theta_t$ as a function of $\delta I_0$ and
$\delta \theta_0$ ($\delta I_0$ and $\delta \theta_0$
represent a change of the initial conditions).
The stability matrix $M$ has eigenvalues
$\mu_{\pm}=\frac{1}{2}(2+K\pm\sqrt{K^2+4K})$, which do not depend
on the coordinates $I$ and $\theta$ and are complex conjugate
for $-4\leq{K}\leq0$ and real for $K<-4$ and $K>0$.
Thus, the classical motion is stable for $-4\leq{K}\leq0$
and completely chaotic for $K<-4$ and $K>0$.
For $K>0$, $\delta_t \propto (\mu_+)^t$ asymptotycally
in $t$, and therefore the maximum Lyapunov exponent
is $\lambda=\ln\mu_+$. Similarly, we obtain
$\lambda=\ln|\mu_-|$ for $K<-4$. In the stable region
$-4\le{K}\le0$, $\lambda=0$.

The sawtooth map can be studied on the cylinder [$I\in(-\infty,+\infty)$],
or on a torus of sinite size ($-{\pi}L\le I < \pi L$, where $L$ is an
integer, to assure that no discontinuities are introduced in the second
equation of (\ref{sawmap2}) when $I$ is taken modulus $2{\pi}L$).
Although the sawtooth map is a deterministic system, for $K>0$ and $K<-4$
the motion along the momentum direction is in practice
indistinguishable from a random walk. Thus, one has normal diffusion in
the action (momentum) variable and the evolution of the distribution
function $f(I,t)$ is governed by a Fokker--Planck equation:
\begin{equation}
  \frac{\partial{f}}{\partial{t}} =
  \frac{\partial}{\partial{I}}
  \left(\frac12 D \frac{\partial{f}}{\partial{I}} \right) .
  \label{fokkerplanck}
\end{equation}
The diffusion coefficient $D$ is defined by
\begin{equation}
  D = \lim_{t\to\infty} \frac{\langle(\Delta{I}_t)^2\rangle}{t} \,,
\end{equation}
where $\Delta{I}\equiv{I}-\langle{I}\rangle$, and $\langle\dots\rangle$ denotes
the average over an ensemble of trajectories. If at time $t=0$ we take
a phase space distribution with initial momentum $I_0$ and random phases
$0\leq\theta<2\pi$, then the solution of the Fokker--Planck equation
(\ref{fokkerplanck}) is given by
\begin{equation}
  f(I,t) =
  \frac1{\sqrt{2 \pi D t}} \, \exp\!\left[ -\frac{(I-I_0)^2}{2Dt} \right] .
\end{equation}
The width $\sqrt{\langle(\Delta I_t)^2\rangle}$
of this Gaussian distribution grows in time, according to
\begin{equation}
 \langle (\Delta{I}_t)^2 \rangle \approx
 D(K) \, t \,.
\end{equation}
For $K>1$, the diffusion coefficient is well approximated by the random phase
approximation, in which we assume that there are no correlations between the
angles (phases) $\theta$ at different times. Hence, we have
\begin{equation}
  D(K) \approx
  \langle   (\Delta{I}^{(1})^2 \rangle =
  \frac1{2\pi}\int_0^{2\pi} d\theta  (\Delta{I}^{(1)})^2  =
  \frac1{2\pi}\int_0^{2\pi} d\theta  K^2(\theta-\pi)^2 =
  \frac{\pi^2}{3}  K^2 ,
\end{equation}
where $\Delta{I}^{(1)}=
{I}_{t+1}-I_t$ is the change in action after a single map step.
For $0<K<1$ diffusion is slowed, due to the sticking of trajectories
close to broken tori (known as cantori), and we have
$D(K)\approx3.3\,K^{5/2}$ (this regime is discussed 
in Ref.~\cite{percival}). For
$-4<K<0$ the motion is stable, the phase space has a complex structure of
elliptic islands down to smaller and smaller scales, and one can observe
anomalous diffusion, that is,
$\langle(\Delta{J})^2\rangle\propto{t}^\alpha$, with $\alpha\ne1$ 
(see Ref.~\cite{benenti01}).
The cases $K=-1,-2,-3$ are integrable.

\subsection{Quantum dynamics}

The quantum version of the sawtooth map
is obtained by means of the usual
quantization rules, $\theta\to\hat{\theta}$ and
$n\to{}\hat{n}=-i\partial/\partial\theta$
(we set $\hbar=1$). The quantum evolution in one map iteration is described by
a unitary operator $\hat{U}$, called the Floquet operator,
acting on the wave vector $|\psi\rangle$:
\begin{equation}
  |\psi\rangle_{t+1} =
  \hat{U}\,|\psi\rangle_t =
  \exp\left[
    -i \int_{lT^-}^{(l+1)T^-} d\tau H(\hat{\theta},\hat{I};\tau)
  \right]
  |\psi\rangle_t \,,
  \label{sawq}
\end{equation}
where $H$ is Hamiltonian (\ref{sawham}). Since the potential $V(\theta)$ is
switched on only at discrete times $lT$, it is straightforward to
obtain
\begin{equation}
  |\psi\rangle_{t+1} =
  e^{-i T \hat{n}^2\!/2} \, e^{-iV(\hat{\theta})} \, |\psi\rangle_t,
  \label{sawquantum}
\end{equation}
which for the sawtooth map is just Eq.~(\ref{eq:quantmap}).
It is important to emphasize that, while the classical
sawtooth map depends only on the rescaled parameter $K=kT$, the
corresponding quantum evolution (\ref{sawquantum}) depends on
$k$ and $T$ separately.
The effective Planck constant is given by $\hbar_{\rm eff}=T$.
Indeed, if we consider the operator $\hat{I}=T\hat{n}$
($\hat{I}$ is the quantization of the classical rescaled action $I$),
we have
\begin{equation}
[\hat{\theta},\hat{I}]=T[\hat{\theta},\hat{n}]=i T =i \hbar_{\rm eff}.
\end{equation}
The classical limit $\hbar_{\rm eff}\to 0$ is obtained by taking
$k\to\infty$ and $T\to0$, while keeping $K=kT$ constant.

In the quantum sawtooth map model one can
observe important physical phenomena like dynamical
localization~\cite{benenti03}. Indeed, due to
quantum interference effects, the chaotic diffusion
in momentum is suppressed, 
leading to exponentially localized wave functions.
This phenomenon was first
found and studied  in the quantum kicked-rotator model~\cite{izrailev} and has
profound analogies with Anderson localization of electronic transport in
disordered materials~\cite{fishman}.
Dynamical localization has been observed experimentally
in the microwave ionization of Rydberg atoms~\cite{koch}
and in experiments with cold atoms~\cite{raizen}.
In the quantum sawtooth map also cantori localization
takes place: In the vicinity of a broken KAM torus,
a cantorus starts to act as a perfect barrier to quantum wave
packet evolution, if the flux through it becomes less
than $\hbar$~\cite{geisel,mackay,fausto,prosen,prange}.

\subsection{Quantum algorithm}

In the following, we describe an exponentially efficient quantum algorithm for
simulation of the map (\ref{eq:quantmap})~\cite{benenti01}.
It is based on the forward/ backward quantum Fourier transform between momentum
and angle bases. Such an approach is convenient since the 
Floquet operator $\hat{U}$,
introduced in Eq.~(\ref{eq:quantmap}), is the product of two operators,
$\hat{U}_k=e^{ik(\hat{\theta}-\pi)^2\!/2}$ and
$\hat{U}_T=e^{-iT\hat{n}^2\!/2}$, diagonal in the $\theta$ and $n$
representations, respectively. This quantum algorithm requires the following
steps for one map iteration:
\begin{itemize}
\item
Apply $\hat{U}_k$ to the wave function $\psi(\theta)$. In order to decompose
the operator $\hat{U}_k$ into one- and two-qubit gates, we first of all
write $\theta$ in binary notation:
\begin{equation}
  \theta=2\pi\sum_{j=1}^{n_q} \alpha_j 2^{-j} \,,
\end{equation}
with $\alpha_i\in \{ 0,1 \}$. Here $n_q$ is the number of qubits, so that the
total number of levels in the quantum sawtooth map is $N=2^{n_q}$. From this
expansion, we obtain
\begin{equation}
  (\theta - \pi)^2 =
  4\pi^2 \sum_{j_1,j_2=1}^{n_q}
  \left( \frac{\alpha_{j_1}}{2^{j_1}} -\frac1{2n} \right)
  \left( \frac{\alpha_{j_2}}{2^{j_2}} -\frac1{2n} \right) .
\end{equation}
This term can be put into the unitary operator 
$\hat{U}_k$, giving the decomposition
\begin{equation}
  e^{ik(\theta -\pi)^2\!/2} =
  \prod_{j_1,j_2=1}^n
  \exp\!\left[
    i 2 \pi^2 k
    \left(\frac{\alpha_{j_1}}{2^{j_1}} - \frac1{2n}\right)
    \left(\frac{\alpha_{j_2}}{2^{j_2}} - \frac1{2n}\right)
  \right] ,
  \label{ukdec}
\end{equation}
which is the product of $n_q^2$ two-qubit gates 
(controlled phase-shift gates), 
each acting non-trivially only on the
qubits $j_1$ and $j_2$. In the computational basis
$\{|\alpha_{j_1}\alpha_{j_2}\rangle =
|00\rangle,|01\rangle,|10\rangle,|11\rangle\}$ each two-qubit gate can be
written as $\exp(i2\pi^2kD_{j_1,j_2})$, where $D_{j_1,j_2}$ is a diagonal
matrix:
\begin{equation}
  D_{j_1,j_2} =
  \left[
    \begin{array}{cccc}
      \frac1{4n^2} & 0 & 0 & 0 \\
      0 & -\frac1{2n}\big(\frac1{2^{j_2}}-\frac1{2n}\big) & 0 & 0 \\
      0 & 0 & -\frac1{2n}\big(\frac1{2^{j_1}}-\frac1{2n}\big) & 0 \\
      0 & 0 & 0 & \big(\frac1{2^{j_1}}-\frac1{2n}\big)
      \big(\frac1{2^{j_2}}-\frac1{2n}\big)
    \end{array}
  \right] .
\end{equation}
\item
The change from the $\theta$ to the $n$ representation is obtained by means of
the quantum Fourier transform, which requires $n_q$ (single-qubit) 
Hadamard gates and $\frac12 n_q(n_q-1)$ 
(two-qubit) controlled phase-shift gates (see, e.g.,~\cite{qcbook,nielsen}).
\item
In the $n$ representation, the operator $\hat{U}_T$ has essentially the same
form as the operator $\hat{U}_k$ in the $\theta$ representation and can
therefore be decomposed into $n_q^2$
controlled phase-shift gates, similarly to Eq.~(\ref{ukdec}).
\item
Return to the initial $\theta$ representation by application
of the inverse quantum Fourier transform.
\end{itemize}
Thus, overall, this quantum algorithm requires $3n_q^2+n_q$
gates per map iteration ($3n_q^2-n_q$ 
controlled phase-shifts and
$2n_q$ Hadamard gates). This number is to be compared with the
$O(n_q2^{n_q})$ operations required by a classical computer to simulate
one map iteration by means of a fast Fourier transform. Thus,
the quantum simulation of the quantum sawtooth map dynamics is exponentially
faster than any known classical algorithm.
Note that the resources required to the quantum computer to simulate
the evolution of the sawtooth map are only logarithmic in the
system size $N$.
Of course, there remains
the problem of extracting useful information from the quantum computer
wave function. For a discussion of this problem, see 
Refs.~\cite{qcbook,georgeotvarenna}.
Finally, I point out that the
quantum sawtooth map has been recently
implemented on a three-qubit nuclear magnetic resonance (NMR)-based
quantum processor~\cite{corysawtooth}. 

\section{Entanglement of formation and concurrence}

\label{sec:concurrence}

Any state $\rho$ can be decomposed as a convex combination of 
projectors onto pure states:
\begin{equation}
\rho=\sum_j p_j |\psi_j\rangle\langle \psi_j|.
\end{equation}
The entanglement of formation $E$ is defined as the mean 
entanglement of the pure states forming $\rho$, minimized over
all possible decompositions:
\begin{equation}
  E(\rho_S) = \inf_{\mathrm{dec}} \sum_j p_j E(|\psi_j\rangle),
\end{equation}
where the (bipartite) entanglement of the pure states $|\psi_j\rangle$
is measured according to Eq.~(\ref{entbipure}).

The entanglement of formation $E_{12}$ of a generic two-qubit state
$\rho_{12}$ can be evaluated in a closed form
following Ref.~\cite{wootters98}.
First of all we compute the \emph{concurrence}, defined as
$C=\max ( \lambda_{1} - \lambda_{2} -
\lambda_{3} - \lambda_{4} , 0 )$,
where the $\lambda_{i}$'s are the square roots of the eigenvalues
of the matrix $R=\rho_{12} \tilde{\rho}_{12}$, in decreasing order.
Here $\tilde{\rho}_{12}$ is the spin flipped matrix of $\rho_{12}$,
and it is defined by $\tilde{\rho}_{12}=
(\sigma_{y} \otimes \sigma_{y}) \, \rho_{12}^{\star} \,
(\sigma_{y} \otimes \sigma_{y})$
(note that the complex conjugate is taken in the computational
basis $\{ \vert 00 \rangle, \vert 01 \rangle, \vert 10 \rangle,
\vert 11 \rangle \}$).
Once the concurrence has been computed, the entanglement
of formation is obtained
as $E= h((1+\sqrt{1-C^{2}})/2)$,
where $h$ is the binary
entropy function: $h(x)=-x\log_{2}x-(1-x)\log_{2}(1-x)$,
with $x=(1+\sqrt{1-C^{2}})/2$.

The concurrence is widely investigated in condensed matter physics,
in relation to the general problem of the behavior of entanglement
across quantum phase transitions~\cite{fazioreview}. For studies
of the relation between entanglement and 
integrability to chaos crossover in quantum spin chain,
see~\cite{simone,monasterio,viola1,viola2,viola3,prosen2007}, 
and references therein. 
 
\acknowledgments
While working on the topics discussed in this review paper, I had 
the pleasure to collaborate with 
Dima Averin, Gabriel Carlo, Giulio Casati, Rosario Fazio,
Giuseppe Gennaro, Jae Weon Lee, 
Carlos Mej\'{\i}a-Monasterio, Simone Montangero, 
Massimo Palma, Toma\v z Prosen, Alessandro Romito,
Davide Rossini, Dima Shepelyansky, Valentin Sokolov,
Oleg Zhirov and Marko \v Znidari\v c.
I would like to express my gratitude to all of them.

\end{document}